\documentstyle[aps,epsfig]{revtex}
\begin{document}
\draft

\def\qb{\bar{q}}
\def\go{\tilde{g}}
\def\sfa{\tilde{f}}
\def\sfb{\overline{\tilde{f}}}
\def\sq{\tilde{q}}
\def\sqb{\bar{\tilde{q}}}
\def\sqlr{\tilde{q}_{\rm L,R}}
\def\stez{\tilde{t}_{1,2}}
\def\sbez{\tilde{b}_{1,2}}
\def\po{\tilde{\gamma}}
\def\zo{\tilde{Z}}
\def\nhoe{\tilde{H}^0_1}
\def\nhoz{\tilde{H}^0_2}
\def\noev{\tilde{\chi}^0_{1-4}}
\def\noi{\tilde{\chi}^0_i}
\def\noe{\tilde{\chi}^0_1}
\def\noz{\tilde{\chi}^0_2}
\def\wo{\tilde{W}^{\pm}}
\def\hoee{\tilde{H}_1^1}
\def\hoez{\tilde{H}_1^2}
\def\hoze{\tilde{H}_2^1}
\def\hozz{\tilde{H}_2^2}
\def\choe{\tilde{H}^{\pm}_1}
\def\choz{\tilde{H}^{\pm}_2}
\def\coez{\tilde{\chi}^{\pm}_{1,2}}
\def\coi{\tilde{\chi}^{\pm}_i}
\def\coe{\tilde{\chi}^{\pm}_1}
\def\coz{\tilde{\chi}^{\pm}_2}
\def\copi{\tilde{\chi}^+_i}
\def\comi{\tilde{\chi}^-_i}
\def\sllr{\tilde{l}_{\rm L,R}}
\def\snl{\tilde{\nu}_{\rm L}}

\def\ms{m_{\tilde q}}
\def\mf{m_{\tilde f}}
\def\mg{m_{\tilde g}}
\def\ghat{\hat{g}_s}
\def\eps{\varepsilon}

 \def\lp{\left. }
 \def\rp{\right. }
 \def\lr{\left( }
 \def\rr{\right) }
 \def\le{\left[ }
 \def\re{\right] }
 \def\lg{\left\{ }
 \def\rg{\right\} }
 \def\lb{\left| }
 \def\rb{\right| }

 \def\met{\rlap{\,/}E_T}

 \def\beq{\begin{equation}}
 \def\eeq{\end{equation}}
 \def\bea{\begin{eqnarray}}
 \def\eea{\end{eqnarray}}

\preprint{\vbox{\halign{&##\hfil\cr& hep-ph/0008081 \cr& DESY 00-112 \cr}}}
\title{Sfermion Pair Production in Polarized and Unpolarized $\gamma\gamma$
 Collisions}
\author{S.\ Berge, M.\ Klasen, Y.\ Umeda}
\address{II.\ Institut f\"ur Theoretische Physik, Universit\"at Hamburg, \\
Luruper Chaussee 149, D-22761 Hamburg, Germany}
\date{\today}
\maketitle
\begin{abstract}
We calculate total and differential cross sections for the production of
sfermion pairs in photon-photon collisions, including contributions from
resolved photons and arbitrary photon polarization. Sfermion production
in photon collisions depends only on the sfermion mass and charge. It
is thus independent of the details of the
SUSY breaking mechanism, but highly sensitive to
the sfermion charge. We compare the total cross sections for bremsstrahlung,
beamstrahlung, and laser backscattering photons to those in $e^+e^-$
annihilation. We find that the total cross section at a polarized photon
collider is larger than the $e^+e^-$ annihilation cross section up to the
kinematic limit of the photon collider.
\end{abstract}
\vspace{0.2in}

\section{Introduction}
\label{sec:1}

\vspace*{-10.3cm}
\noindent hep-ph/0008081 \\
DESY 00-112
\vspace*{9.8cm}

Among the possible extensions of the Standard Model (SM), supersymmetric (SUSY)
theories have a variety of attractive features: They can solve the Higgs
hierarchy puzzle, break electroweak symmetry radiatively at low energies, and
explain the unification of gauge couplings at a high energy scale. A
necessary condition for these arguments to be valid is that SUSY is realized
in the region of the electroweak scale. This makes the search for supersymmetry
one of the
most important tasks at future high-energy collider experiments.

If SUSY is realized at the electroweak scale, most of the SUSY partners
of the SM particles will be discovered at the high center-of-mass energies
available at the next generation of hadron colliders, {\it i.e.} at Run II
of the Fermilab Tevatron or at the CERN LHC. After this discovery stage
it will be important to analyze the properties of these sparticles and to
check whether they have the correct quantum numbers to be partners of the
SM particles.

While hadron colliders have the advantage of large available center-of-mass
energy, they also have serious disadvantages: First they produce enormous
backgrounds from SM processes making it difficult to distinguish the signal
from the background. Second, the remnants of the initial hadrons make it
impossible to reconstruct the full final state. Third, the energies of the
partons initiating the hard scattering are unknown so that an energy (mass)
scan becomes impossible. 

In $e^+e^-$ annihilation, the full center-of-mass energy participates in the
hard scattering and is precisely known, and the final state consists of a
small number of high-energetic particles. The precision studies following
upon the SUSY discovery stage are therefore the natural domain of high-energy
lepton colliders. Important information on the SUSY parameters can be
gained from the SUSY mass spectrum: It will be important to know
whether all sfermions, squarks and sleptons, or those of the same
generation have identical mass parameters and/or gauge couplings
or how large the differences
and ratios among them are, or how much the left- and right-handed squarks
or sleptons are mixed \cite{Accomando:1998wt}.

Currently several linear $e^+e^-$ colliders in the 500-3000 GeV center-of-mass
energy
range are under design in various international collaborations. From previous
experience with existing lepton colliders like LEP2 it is well-known that
photons will be ubiquitous at future lepton colliders due to bremsstrahlung
and beamstrahlung effects. Furthermore it has been proposed to backscatter
laser photons from the lepton beams in order to build a
collider with high-energetic and almost monochromatic photon beams
\cite{Ginzburg:1984yr}. Photon colliders have similar advantages as lepton
colliders: In energy scans the initial photon energy is known, although
only to $\pm 15\%$, and the final state can be completely reconstructed.

Sfermion production in photon collisions has been considered previously
either with bremsstrahlung photons \cite{Grifols:1983gs,Rizzo:1989ip} or with
laser photons, where the center-of-mass energy dependence for fixed
sfermion mass was analyzed in \cite{Cuypers:1993tj,Chang:1998tw} and the
sfermion mass dependence for a fixed collider energy of 1 TeV in
\cite{Telnov:2000tb,{Choudhury:2000ue}}. The production of bound squarkonium
states was considered in \cite{Bigi:1991vf} for bremsstrahlung photons and in
\cite{Gorbunov:2000tr} for photon colliders.

In this paper we present the first complete analysis of sfermion production
in photon-photon collisions from bremsstrahlung, beamstrahlung, and laser
backscattering, including resolved photon processes and polarization
effects. We compare the total cross sections directly to those in $e^+e^-$
annihilation and also present differential cross sections.
A FORTRAN program to generate total or differential cross sections for any
sfermion type in polarized or unpolarized photon-photon collisions or in
$e^+e^-$ annihilation is available from the authors upon request.

In Sec.\ \ref{sec:2} we review for completeness the unpolarized photon spectra
coming from bremsstrahlung, beamstrahlung, and laser backscattering and update
them using the latest linear collider design parameters.
In Sec.\ \ref{sec:3} we present our analytical and numerical results for
sfermion production in unpolarized photon-photon collisions and compare
them to those in $e^+e^-$ annihilation.
Sec.\ \ref{sec:4} contains a discussion of polarized photon spectra.
In Sec.\ \ref{sec:5} we calculate analytically and numerically total and
differential cross sections for sfermion production in polarized photon-photon
collisions. 
Our conclusions are given in Sec.\ \ref{sec:6}.


\section{Unpolarized Photon Spectra}
\label{sec:2}

High energy electron-positron colliders are abundant sources of photons
due to the presence of three photon production mechanisms:
bremsstrahlung, beamstrahlung, and laser backscattering. While the first
two radiation processes occur at any circular or linear $e^+e^-$
collider, albeit at different levels, laser backscattering requires
additional laser beams and focusing mirrors, which may
also interfere with the design of the detectors. These modifications
still pose technical difficulties, and they will also increase the cost
of such a ``photon collider''.

Bremsstrahlung can be conveniently described through an approximation of
the complete two-photon process $e^+e^-\rightarrow e^+e^-X$. The
outgoing photon spectrum is given by the Weizs\"acker-Williams formula
\cite{deFlorian:1999ge}
\beq
 f_{\gamma/e}^{\rm brems}(x)=\frac{\alpha}{2\pi}\left[
 \frac{1+(1-x)^2}{x}\ln\frac{Q^2_{\max}(1-x)}{m_e^2 x^2}
 +2 m_e^2 x\left(\frac{1}{Q^2_{\max}}-\frac{1-x}{m_e^2 x^2}\right)\right].
 \label{eq:unpol_brems}
\eeq
It has been integrated over the photon virtuality up to an upper bound
$Q^2_{\max}=4E_e^2(1-x)$ for untagged outgoing electrons, which depends on the
electron (positron) beam energy $E_e=\sqrt{S}/2$. This leads to a logarithmic
dependence of the spectrum on $S$, the squared center-of-mass energy of the
collider. $\alpha=e^2/(4\pi)=\sqrt{2}G_Fm_W^2s_W^2/\pi$ is the electromagnetic
coupling constant in the $G_F$ scheme,
where $G_F$ is the Fermi coupling constant, $s_W = \sqrt{1-(m_W/m_Z)^2}$
is the sine of the electroweak mixing angle, and $m_Z = 91.187$ GeV and
$m_W = 80.41$ GeV are the masses of the electroweak gauge bosons.
$m_e$ is the electron mass, and $x$ is the fractional energy of the photon
in the electron.

At existing electron-positron or electron-proton colliders like LEP2 and
HERA, bremsstrahlung is the only relevant source of photons. Future
circular electron-positron colliders above $\sqrt{S}=500$ GeV would
suffer from very high synchrotron radiation. They must therefore have a linear
design with large luminosities and dense particle bunches. Inside the opposite
bunch, electrons and positrons experience transverse acceleration and radiate
beamstrahlung. The corresponding spectrum \cite{Schroeder:1990bx}
\beq
 f_{\gamma/e}^{\rm beam}(x) = \frac{5}{4\sqrt{3}\Upsilon}
 \int_u^{\infty}{\rm d}v
 {\rm Ai}(v)\left[\left(\frac{2v}{u}-1\right)\frac{1+(1-x)^2}{2(1-x)}
 +\frac{x^2}{2(1-x)}\right],
 \label{eq:unpol_beam}
\eeq
\begin{table}
\begin{tabular}{|c|cccc|}
  Collider                        & TESLA& JLC  & NLC   & CLIC \\
  Last update                     & 8/98 & 9/99 & 12/98 & 9/99 \\
\hline
\hline
  Center-of-mass energy (GeV)     & 500  & 500  & 500   & 500  \\
  Particles per Bunch ($10^{10}$) & 2    & 1.11 & 0.95  & 0.4  \\
  $\sigma_x$ (nm)                 & 553  & 318  & 330   & 202  \\
  $\sigma_y$ (nm)                 & 5    & 4.3  & 4.9   & 2.5  \\
  $\sigma_z$ ($\mu$m)             & 400  & 200  & 120   & 30   \\
  $\Upsilon$                      & 0.038& 0.074& 0.101 & 0.280\\
\hline
  Center-of-mass energy (GeV)     & 800  & 1000 & 1000  & 1000 \\
  Particles per Bunch ($10^{10}$) & 1.41 & 1.39 & 0.95  & 0.4  \\
  $\sigma_x$ (nm)                 & 391  & 318  & 234   & 115  \\
  $\sigma_y$ (nm)                 & 2    & 3.14 & 3.9   & 1.75 \\
  $\sigma_z$ ($\mu$m)             & 300  & 200  & 120   & 30   \\
  $\Upsilon$                      & 0.082& 0.186& 0.285 & 0.979
\end{tabular}
\caption{\label{tab:colliders}Current design parameters for possible future
 linear colliders.}
\end{table}
\begin{figure}
 \begin{center}
  \epsfig{file=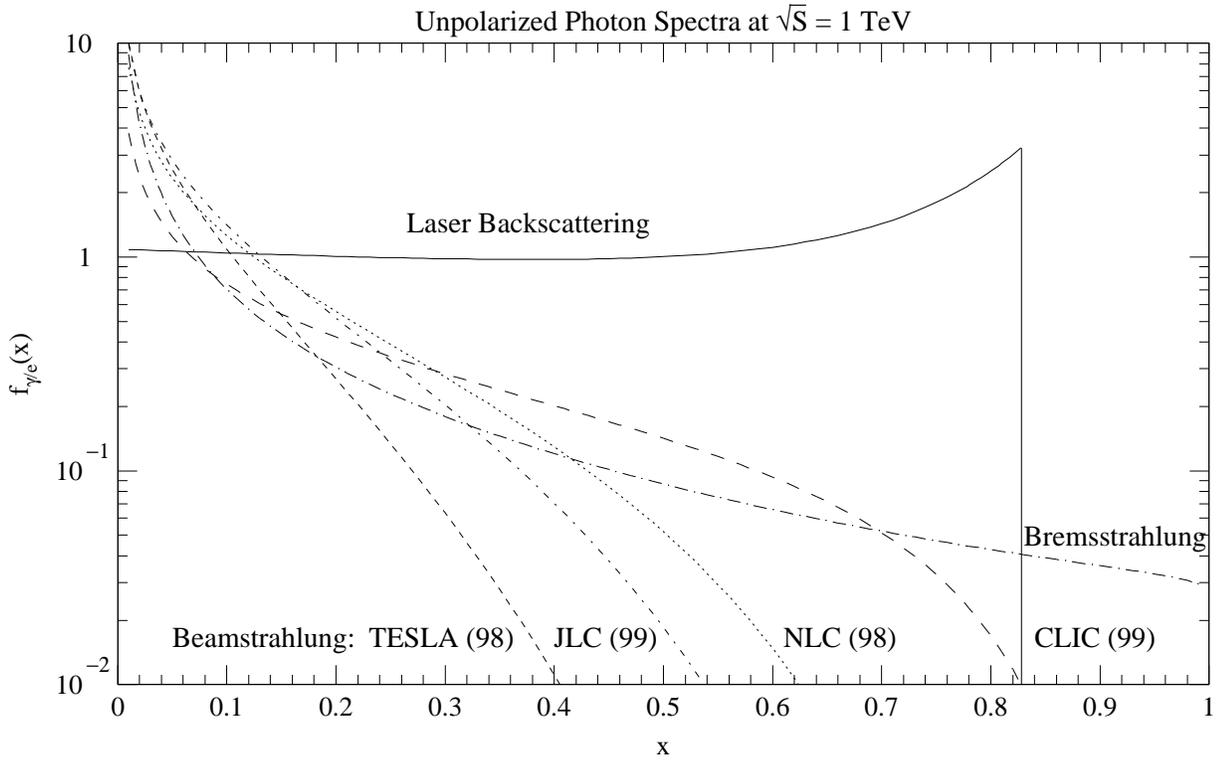,width=\textwidth,clip=}
 \end{center}
 \caption{\label{fig:phospec_unpol}
 Photon spectra for $\sqrt{S} = 1$ TeV $e^+e^-$ colliders. For TESLA, the
 1998 beam size parameters of the $\sqrt{S} = 0.8$ TeV design have been used.}
\end{figure}
\noindent
where the Airy-function Ai$(v)$ falls off exponentially at large $v$ and
$u=\{5x/[4\sqrt{3}\Upsilon(1-x)]\}^{2/3}$, is controlled by the beamstrahlung parameter
\beq
 \Upsilon = \frac{5 r_e^2 E_e N}{6 \alpha \sigma_z (\sigma_x+\sigma_y) m_e}.
\eeq
This parameter is proportional to the effective electromagnetic field of the bunches
and depends on the classical electron radius $r_e = \alpha/m_e = 2.818
\cdot 10^{-15}$ m, on the r.m.s.\ sizes of the Gaussian beam $\sigma_x,~
\sigma_y,~\sigma_z$, and on the total number of particles in a bunch
$N$. Current design parameters for future linear colliders are listed in Tab.\
\ref{tab:colliders}\cite{ref:colliders}.

For not too large $\Upsilon \leq 5$ the spectrum for multiple photon emission
can be written in the approximate form \cite{Chen:1992wd}
\bea
\label{eq:appbeam}
 f_{\gamma/e}^{\rm beam}(x) &=& \frac{1}{\Gamma\left(\frac{1}{3}\right)}
  \left(\frac{2}{3\Upsilon}\right)^\frac{1}{3} x^{-\frac{2}{3}}
  (1-x)^{-\frac{1}{3}} \exp\left[-\frac{2 x}{3\Upsilon (1-x)}\right] \\
 & \times & \left\{\frac{1-\sqrt{\frac{\Upsilon}{24}}}{g(x)}\left[ 1-\frac{1}
  {g(x) N_\gamma}\left(1-e^{-g(x) N_\gamma}\right)\right]
  +\sqrt{\frac{\Upsilon}{24}} \left[1-\frac{1}{N_\gamma} \left( 
  1-e^{-N_\gamma}\right)\right]\right\}, \nonumber
\eea
where
\beq
 g(x) = 1 - \frac{1}{2} \left[(1+x)\sqrt{1+\Upsilon^{\frac{2}{3}}}+1-x\right]
 (1-x)^{\frac{2}{3}}
\eeq
and the average number of photons radiated per electron throughout the
collision is
\beq
 N_\gamma = \frac{5 \alpha^2 \sigma_z m_e}{2 r_e E_e} \frac{\Upsilon}
 {\sqrt{1+\Upsilon^{\frac{2}{3}}}}.
\eeq
Since $\Upsilon\propto\sqrt{S}$, the exponential suppression of beamstrahlung
decreases with rising $\sqrt{S}$ and from up to down in Tab.~\ref{tab:colliders}.
Beamstrahlung is most important for the CLIC design
and less important for the NLC, JLC, or TESLA designs.

The photon spectra for the $\sqrt{S} = 1$ TeV $e^+e^-$ colliders listed in
Tab.\ \ref{tab:colliders} are shown in Fig.\ \ref{fig:phospec_unpol},
where we have used the beamstrahlung spectrum of Eq.\ (\ref{eq:appbeam}).
It is obvious that the bremsstrahlung and beamstrahlung mechanisms produce
mostly soft photons. However, it is interesting to note that beamstrahlung at
a 3 TeV CLIC collider \cite{ref:clic} produces already three times more
high-energetic photons (at $x=0.8$) than bremsstrahlung alone.

The situation is much better if laser photons are backscattered off the
incident lepton beams. The laser backscattering spectrum \cite{Ginzburg:1984yr}
\beq
 f_{\gamma/e}^{\rm laser}(x) = \frac{1}{N_c} \left[1-x+\frac{1}{1-x}-\frac{4 x}
 {X (1-x)}+\frac{4 x^2}{X^2 (1-x)^2}\right] ,
\eeq
where
\beq
 N_c = \left[1-\frac{4}{X}-\frac{8}{X^2}\right] \ln(1+X)+\frac{1}{2}
 + \frac{8}{X}-\frac{1}{2 (1+X)^2}
\eeq
is related to the total Compton cross section, depends on the center-of-mass
energy of the electron-laser photon collision $X = 4 E_e E_\gamma/m_e^2$. The
optimal value of $X$ is determined by the threshold for the process
$\gamma\gamma\rightarrow e^+e^-$ and is $X=2+\sqrt{8}\simeq 4.83$
\cite{Telnov:1990sd}. If this value is kept fixed, the laser backscattering
spectrum becomes independent of $\sqrt{S}$.  A large fraction of the photons
is then produced close to the kinematic limit $x < x_{\max} = X/(1+X)
\simeq 0.828$, so that one obtains an almost monochromatic ``photon collider''.


\section{Unpolarized Cross Sections}
\label{sec:3}

The inclusive cross section for photoproduction of sfermions in
electron-positron collisions
\beq
 \sigma_{e^+e^-}^B(S) = \int {\rm d}x_1 f_{i/e}(x_1) {\rm d}x_2 f_{j/e}(x_2)
 {\rm d}t_{\sfa}{\rm d}u_{\sfa}\frac{{\rm d}^2\sigma^B_{ij}(s)}
 {{\rm d}t_{\sfa}\,{\rm d}u_{\sfa}}
 \label{eq:ee_xsec_unpol}
\eeq
can be obtained by convolving the hard photonic cross section
\beq
 \frac{{\rm d}^2\sigma^B_{ij}(s)}{{\rm d}t_{\sfa}\,{\rm d}u_{\sfa}} =
 \frac{\pi (4\pi)^{-2+\eps}}{s^2\,\Gamma (1 - \eps)} \left[\frac{ t_{\sfa}\,u_{\sfa}
 - \mf^2 s}{\mu^2\,s} \right]^{- \eps} \Theta(t_{\sfa}\,u_{\sfa} -m^{2}_{\sfa}
 \,s) \Theta (s - 4 m_{\sfa}^2)\, \delta (s + t_{\sfa} + u_{\sfa})\,
 \overline{|{\cal M}^B_{ij}|}^{2}
 \label{eq:yy_xsec_unpol}
\eeq
with the photon energy spectra $f_{\gamma/e}(x)$ discussed in the
previous Section. We denote the momenta of the massless incoming photons by
$k_{1,2}$ and those of the outgoing sfermions with mass $\mf$ by $p_{1,2}$.
$s=(k_1+k_2)^2=x_1x_2S,~t_{\sfa}=(k_2-p_2)^2-\mf^2,$ and
$u_{\sfa}=(k_1-p_2)^2-\mf^2$
are the Mandelstam variables of the hard photon-photon scattering process.
$\overline{|{\cal M}^B_{ij}|}^{2}$ is the
partonic matrix element squared, summed (averaged) over final (initial) state
spins and calculated in $d=4-2\eps$ dimensions. $\mu$ is an arbitrary
scale parameter.
In addition to the direct contributions with $i,j=\gamma$, one
or two of the photons can also resolve into a hadronic structure before they
interact. For these single- and double-resolved contributions, the photon
spectra $f_{\gamma/e}(x)$ have to be convolved with the parton density
functions $f_{i,j/\gamma} (y)$ of quarks or gluons in the photons
\beq
 f_{i,j/e}(x) = \int_x^1 \frac{{\rm d}y}{y} f_{\gamma/e}\left(\frac{x}{y}
 \right) f_{i,j/\gamma} (y).
\eeq
\begin{figure}
 \begin{center}
  \epsfig{file=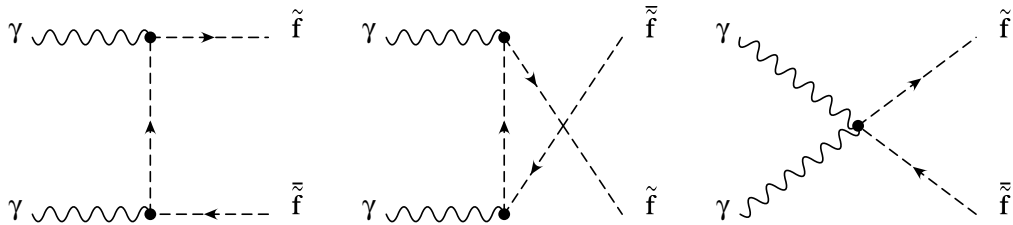,bbllx=62pt,bblly=360pt,bburx=373pt,bbury=434pt,%
          width=13.5cm,clip=}
 \end{center}
 \caption{\label{fig:dlo}
 Leading order Feynman diagrams for direct sfermion production in
 photon-photon collisions.}
\end{figure}

\begin{figure}
 \begin{center}
  \epsfig{file=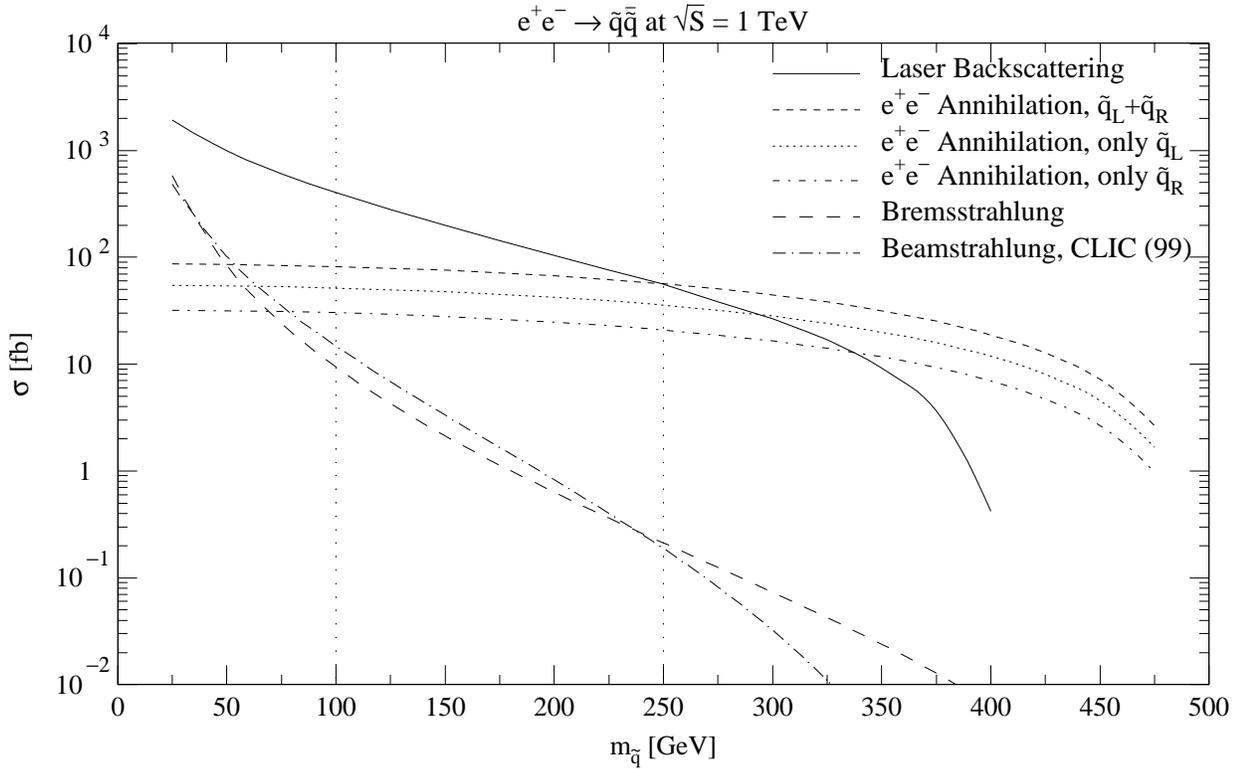,width=\textwidth,clip=}
 \end{center}
 \caption{\label{fig:sigma_ms}
 Total cross sections for up-type squark ($\sq_L+\sq_R$) production in
 photon-photon scattering
 and in $e^+e^-$ annihilation at a 1 TeV collider as a function of the squark
 mass.}
\end{figure}
\noindent
The parton densities in the photon can not be calculated in perturbation theory
but have to be fitted to experimental data, e.g.\ on the photon structure
function $F_2^\gamma(x)$ or on jet photoproduction.

In contrast, the partonic matrix elements can be calculated in perturbation
theory. Direct sfermion production in leading order proceeds through the
three diagrams shown in Fig.\ \ref{fig:dlo}.
The corresponding matrix element is given by
\beq
 \overline{|{\cal M}^B_{\gamma\gamma}|}^2=\frac{4 e^4 e_{\sfa}^4 N_C
 \left[\left( 1 - \eps \right) t_{\sfa}^2 u_{\sfa}^2 - 2 \mf^2 t_{\sfa}
 u_{\sfa} s + 2 \mf^4 s^2\right]}{(1-\eps)^2 t_{\sfa}^2 u_{\sfa}^2} .
\eeq
Here, summing over left- and right-handed squarks and sleptons has led to an
additional factor of 2, but we do not sum over different sfermion
generations. The color factor $N_C=3$ for squarks, and $N_C=1$ for
sleptons. Due to
the large mass of the top quark, the supersymmetric partners of the left- and
right-handed top, $\tilde{t}_{L,R}$, can mix to $\tilde{t}_{1,2}$ as can those
of the bottom quark.
It is important to note that the direct photon-photon
cross section is proportional to the fourth power of the sfermion charge
$ee_{\sfa}$ $(e_{\tilde{u}}= 2/3,~e_{\tilde{d}}=-1/3,~
e_{\tilde{\ell}}=-1)$. The cross
section is independent of the details of the
SUSY breaking mechanism, since it depends
only on the physical sfermion masses $\mf$.

This is in contrast to electron-positron annihilation, where the squared
matrix element
\bea
 \overline{|{\cal M}^B_{e^+e^-}|}^2 &=& \frac{2 \delta_{ij}e^4 e_{\sfa}^2 N_C 
 \left( t_{\sfa} u_{\sfa}-\mf^2 s \right) }{s^2}
  \label{eq:sigma_ee} \\
  &+& \frac{a_{ij}^2 e^4 N_C 
     \left( 1 - 4 s_W^2 + 8 s_W^4 \right)  
     \left( t_{\sfa} u_{\sfa} - \mf^2 s \right) }{64 s_W^4 c_W^4 
     \left( s^2 + m_Z^4 + 
       m_Z^2 \left( -2 s + \Gamma_Z^2 \right)  \right) 
     } \nonumber \\
  &+& \frac{a_{ij}\delta_{ij} e^4 e_{\sfa} N_C 
     \left( s - m_Z^2  \right)  
     \left( 1 - 4 s_W^2 \right)  
     \left( t_{\sfa} u_{\sfa} - \mf^2 s \right) }{4 s_W^2 c_W^2 s
     \left( s^2 + m_Z^4 + 
       m_Z^2 \left( -2 s + \Gamma_Z^2 \right)  \right) 
     } \nonumber
\eea
\begin{figure}
 \begin{center}
  \epsfig{file=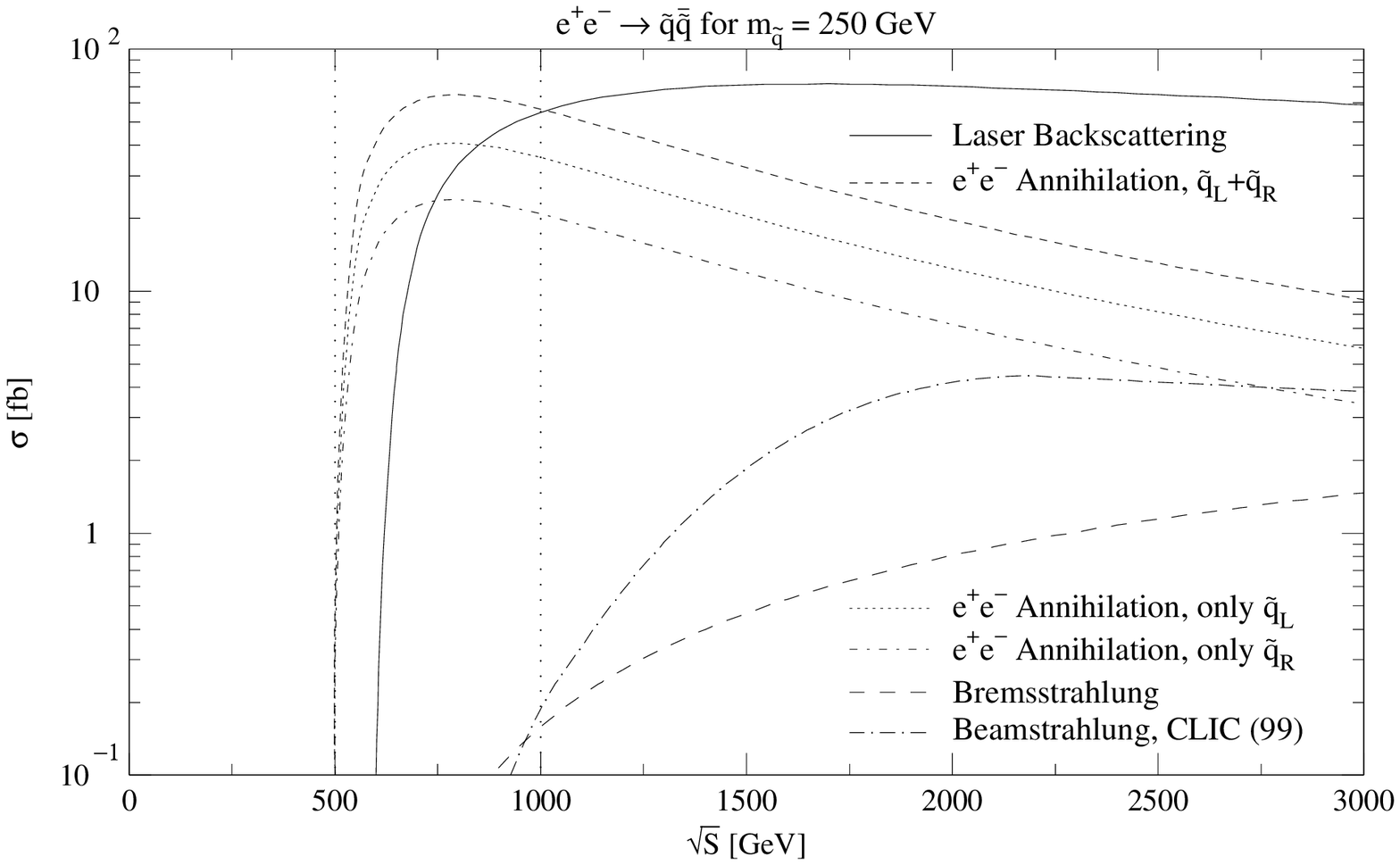,width=\textwidth,clip=}
 \end{center}
 \caption{\label{fig:sigma_250_s}
 Total cross sections for the production of
 up-type squarks ($\sq_L+\sq_R$) of mass 250 GeV in
 photon-photon scattering and in $e^+e^-$ annihilation as a function of
 the center-of-mass energy.}
\end{figure}
\noindent
depends on the details of the
SUSY breaking mechanism through the sfermion mixing angle
$\theta_{\sfa}$ in the $Z^0\sfa_i\sfb_j$ coupling $e/(4 s_W c_W) a_{ij}$
with
\beq
a_{ij} = \left( \begin{array}{cc}
4 ( I_f^{3L} \cos^2\theta_{\sfa} -  e_{\sfa} s_W^2) &
-2 I_f^{3L} \sin 2\theta_{\sfa}\\
-2 I_f^{3L} \sin 2\theta_{\sfa} &
 4 ( I_f^{3L} \sin^2\theta_{\sfa} -  e_{\sfa} s_W^2)
\end{array} \right) ,
\eeq
and it is possible to produce off-diagonal sfermion mass eigenstates
$\sfa_i\sfb_j$.
In this paper we restrict ourselves to the case of no squark mixing.
$s_W (c_W)$ is the sine (cosine) of the electroweak mixing angle
$\theta_W$ and $I_f^{3L} = 1/2\, (-1/2)$ for up-type (down-type) fermions $f$.
The first term in Eq.\ (\ref{eq:sigma_ee})
corresponds to the exchange of a photon in the $s$-channel, the second one to
the exchange of a $Z^0$ boson, and the third one to
the interference between the two.
The cross section for the production of
sfermions in electron-positron annihilation
\beq
 \sigma_{e^+e^-}^B(S) = \int 
 {\rm d}t_{\sfa}{\rm d}u_{\sfa}\frac{{\rm d}^2\sigma^B_{e^+e^-}(S)}
 {{\rm d}t_{\sfa}\,{\rm d}u_{\sfa}}
\eeq
does of course not depend on photon spectra or parton densities.

In Fig.\ \ref{fig:sigma_ms} we show the total cross section for up-type
squark
production in photon-photon scattering and in $e^+e^-$ annihilation at
$\sqrt{S}= 1$ TeV as a
function of the squark mass. Only the annihilation cross section differs for
left- and right-handed squarks. It extends out to $\ms \leq \sqrt{S}/2$. With laser
backscattering, the mass range is reduced to $\ms \leq 0.8 \sqrt{S}/2$.
However, the photoproduction cross section exceeds the annihilation cross
section for $\ms \leq 250$ GeV by up to an order of magnitude, and even above
250 GeV it is of comparable size. While up-type squarks below $250$ GeV are
already excluded by current Tevatron data, a light stop and sleptons with
$m_{\tilde{t}_1,\tilde{\ell}}>100$ GeV are still allowed
\cite{Groom:2000in,Cho:2000km,Alam:2000cs}.
These sparticles can therefore better be studied at a photon
collider. Note that the photon cross section for  
\begin{figure}
 \begin{center}
  \epsfig{file=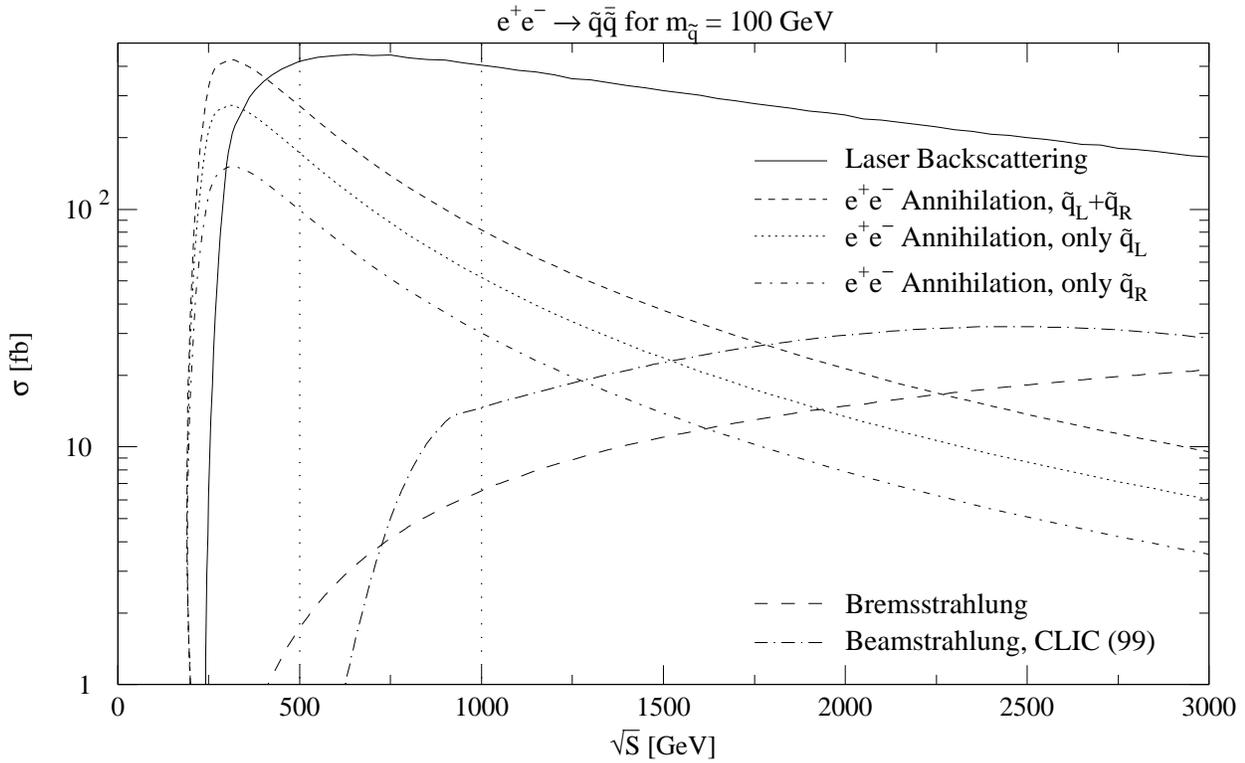,width=\textwidth,clip=}
 \end{center}
 \caption{\label{fig:sigma_100_s}
 Total cross sections for the production of
 up-type ($\sq_L+\sq_R$) squarks of mass 100 GeV in
 photon-photon scattering and in $e^+e^-$ annihilation as a function of
 the center-of-mass energy.}
\end{figure}
\noindent
light left- (right-) handed
sfermions has to be divided by 2 and compared to the left- (right-) handed
annihilation
cross section. Slepton cross sections are larger than up-type squark cross sections
by a factor $1/(3e_{\tilde{u}}^4) = 27/16$, while
down-squark cross sections
are smaller by a factor $(e_{\tilde{d}}/e_{\tilde{u}})^4 = 1/16$.
In addition, selectron production in $e^+e^-$ annihilation is enhanced by
the $t$-channel exchange of a neutralino. In Fig.\ \ref{fig:sigma_ms} we also
show the cross sections for photons produced with brems- and beamstrahlung
without additional laser facilities. We find that the bremsstrahlung cross
section is of similar size as $e^+e^-$ annihilation only for very light
squarks $\ms \leq 100$ GeV, which are already excluded experimentally. The
bremsstrahlung cross section is always lower than the laser cross section
by one to three orders of magnitude. Beamstrahlung, which is most important for
the CLIC (99) design, behaves similarly to bremsstrahlung and enhances the
bremsstrahlung cross section by a factor of 2.\footnote{Our results agree
with \cite{Rizzo:1989ip} for $e^+e^-$ annihilation, but differ for
the bremsstrahlung cross sections by approximately a factor of 3.}

In Fig.\ \ref{fig:sigma_250_s} we show the cross sections for the production
of up-type squarks of mass 250 GeV as a function of the center-of-mass energy
of the collider. The annihilation cross section has a maximum
at $\sqrt{S} \sim 3 \ms$.
At the kinematic limit of the collider, sfermions can only be produced through
the annihilation process, but the laser backscattering cross section also shows
a very steep threshold behavior. At higher energies, the laser backscattering
cross section
stays constant, while the annihilation cross section falls off like $1/S$.
At 1 TeV or higher center-of-mass energy a photon collider is therefore
favorable. At a very large energy CLIC collider even the brems- and
beamstrahlung cross sections become comparable to the annihilation cross
section. For the beamstrahlung cross section it is necessary to interpolate
between $\sqrt{S}=500,~1000,$ and 3000 GeV, since the CLIC (99)
design parameters are only known for these fixed center-of-mass energies.
 
For a light stop and sleptons of mass $m_{\tilde{t}_1,\tilde{\ell}}\simeq 100$
GeV a photon collider is already favorable at the threshold of the
\begin{figure}
 \begin{center}
  \epsfig{file=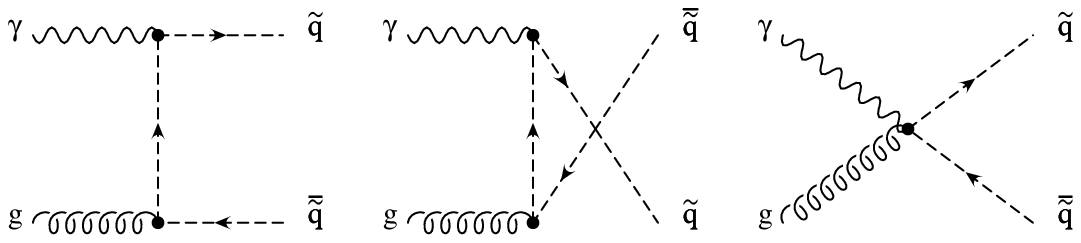,bbllx=62pt,bblly=360pt,bburx=373pt,bbury=432pt,%
          width=13.5cm,clip=}
  \epsfig{file=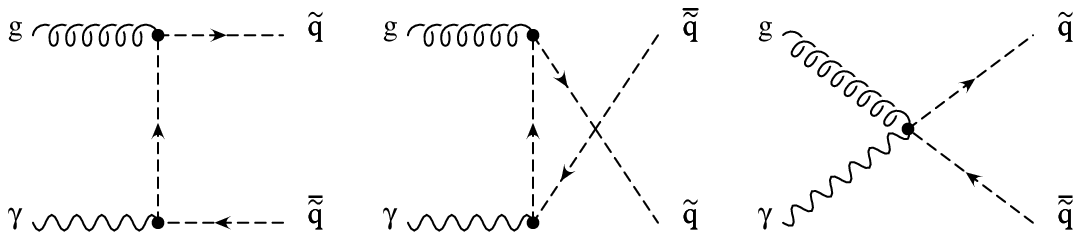,bbllx=62pt,bblly=360pt,bburx=373pt,bbury=432pt,%
          width=13.5cm,clip=}
 \end{center}
 \caption{\label{fig:srlo}
 Leading order Feynman diagrams for single-resolved sfermion production in
 photon-photon collisions. Initial state quarks contribute only at
 next-to-leading order.}
\end{figure}
\begin{figure}
 \begin{center}
  \epsfig{file=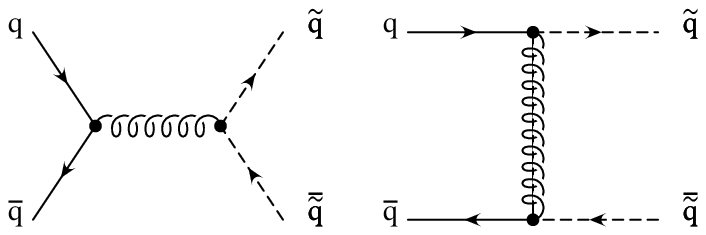,bbllx=62pt,bblly=360pt,bburx=265pt,bbury=432pt,%
          width=9cm,clip=}
  \epsfig{file=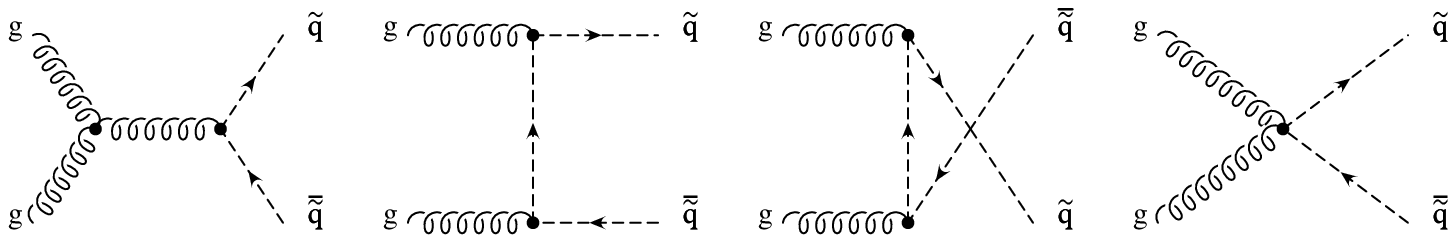,bbllx=62pt,bblly=360pt,bburx=481pt,bbury=432pt,%
          width=\textwidth,clip=}
 \end{center}
 \caption{\label{fig:drlo}
 Leading order Feynman diagrams for double-resolved sfermion production in
 photon-photon collisions.}
\end{figure}
\noindent
sfermion
pair production process in photon-photon collisions ($2 \times 100$ GeV / 0.8
= 250 GeV), and the cross section falls off very slowly at large energies.
This can be seen from Fig.\ \ref{fig:sigma_100_s} where we show total
cross sections for the
production of up-type squarks of mass 100 GeV.
Even the brems- and beamstrahlung cross section for these light sparticles
becomes interesting around $\sqrt{S}=1$ TeV.

As mentioned at the beginning of this Section, photons can not only produce
sfermions by direct coupling, but also after resolving into a hadronic
structure. The direct, single, and double-resolved hard scattering matrix
elements are formally of ${\cal O}(\alpha^2),~{\cal O}(\alpha\alpha_s)$, and
${\cal O}(\alpha_s^2)$, where $\alpha$ and $\alpha_s$ are the electromagnetic
and strong coupling constants. However, the resolved processes have to be
convolved with parton densities in the photon, which are of
${\cal O}(\alpha/\alpha_s)$ at asymptotically large factorization scales
$\mu=\ms$. Therefore all three categories end up being of the same order
${\cal O}(\alpha^2)$ in the perturbative expansion.
In resolved processes, only part of the initial photon energy participates in
the hard scattering. Therefore one expects them to be important only at low
masses, where not the full center-of-mass energy is needed.

If one of the two photons still couples directly, we obtain the contributions
shown in Fig.\ \ref{fig:srlo}. The matrix element squared for the process
$\gamma g\to\sq\sqb$
\beq
 \overline{|{\cal M}^B_{\gamma g}|}^2 = \frac{e^2 e_{\sq}^2 g_s^2 C_F
 \left[\left( 1-\eps \right) t_{\sq}^2 u_{\sq}^2 -2 \ms^2 t_{\sq}u_{\sq}s+
 2 \ms^4 s^2\right]}{2 (1-\eps)^2 N_C C_Ft_{\sq}^2 u_{\sq}^2} \nonumber
\eeq
can be obtained from the direct contribution by the replacements
$e^2e_{\sfa}^2\rightarrow g_s^2$ and $N_C\rightarrow C_F/(2N_CC_F)$,
where 
\begin{figure}
 \begin{center}
  \epsfig{file=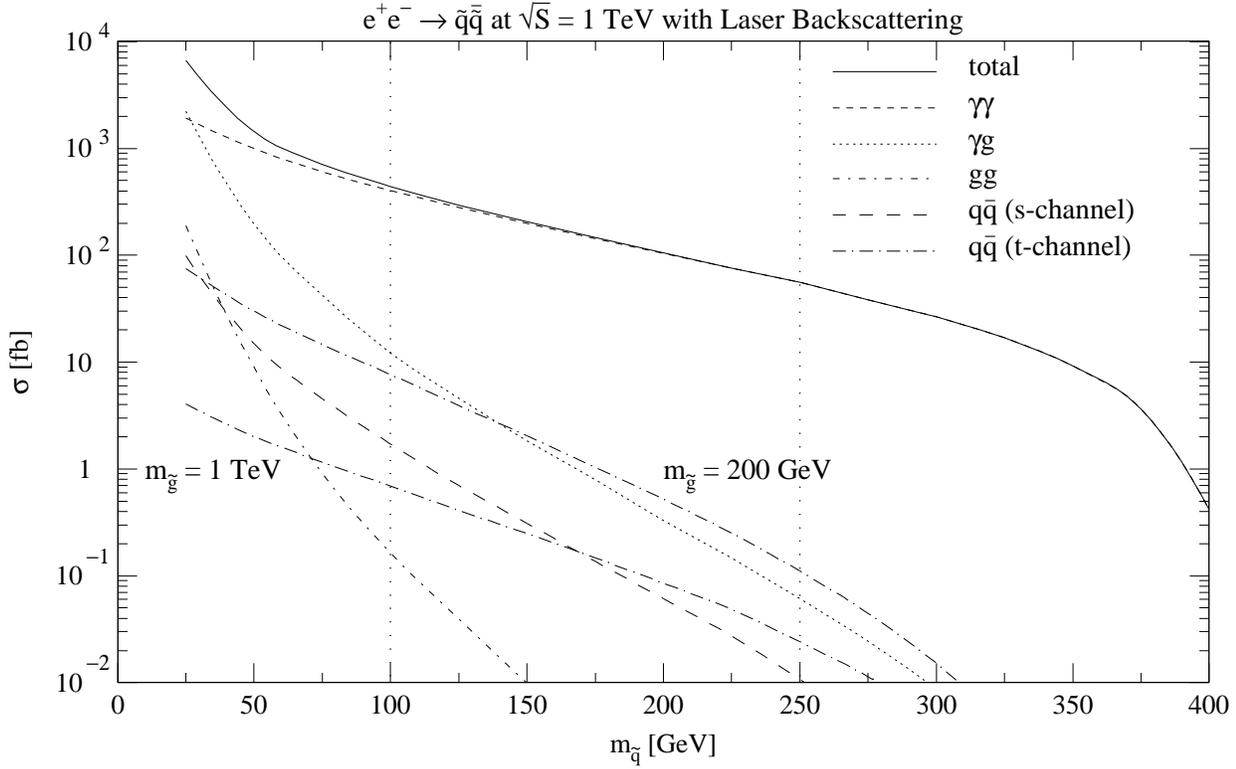,width=\textwidth,clip=}
 \end{center}
 \caption{\label{fig:sigma_msr}
 Total, direct, and resolved cross sections for
 up-type squark ($\sq_L+\sq_R$) production at
 a 1 TeV photon collider as a function of the squark mass.}
\end{figure}
\noindent
$g_s^2/(4\pi) = \alpha_s(\mu)$ is the strong coupling constant and
$C_F=(N_C^2-1)/(2N_C)$ is the color factor of SU(3). The matrix element
squared for the process $g \gamma\to\sq\sqb$ is identical.
Obviously resolved contributions arise only
for squarks, since
sleptons do not couple strongly. The single-resolved
matrix elements have to be convolved with the gluon density in the photon,
which constitutes a higher order ${\cal O}(\alpha/\alpha_s\times\alpha_s)$
contribution to
the photon structure and is not well constrained from
deep-inelastic $e\gamma$ scattering. The gluon density is expected to be small
at large $x=2\ms/\sqrt{s}$.

If both photons resolve into a hadronic structure, we obtain the contributions
in Fig.\ \ref{fig:drlo}. The matrix elements squared for the processes
$q_i\qb_j\rightarrow\sq\sqb$ and $gg\rightarrow\sq\sqb$ are
\bea
 \overline{|{\cal M}^B_{q_i\qb_j}|}^2  & = & 
 \frac{\delta_{ij}}{4 N_C^2}\left[8 g_s^4 N_C C_F \frac{t_{\sq} u_{\sq}
 -\ms^2 s}{s^2} + 4 \ghat^4  N_C C_F 
  \frac{t_{\sq} u_{\sq} -(\ms^2 -\mg^2) s}{(t-\mg^2)^2}
  - 8 g_s^2 \ghat^2  C_F  \frac{t_{\sq} u_{\sq} -\ms^2 s}{s (t-\mg^2)}
 \right] \\ &+& \frac{1-\delta_{ij}}{4 N_C^2}
  \left[ 4 \ghat^4  N_C C_F  \frac{t_{\sq} u_{\sq} -(\ms^2 -\mg^2)s}
 {(t-\mg^2)^2}\right], \nonumber\\
 \overline{|{\cal M}^B_{gg}|}^2 & = & \frac{4 g_s^4}{16 (1-\eps)^2 N_C^2 C_F^2}
 \left[C_O\left(1 -2 \frac{t_{\sq} u_{\sq}}{s^2}\right) -C_K\right]
  \left[ 1 -\eps -2 \frac{s\ms^2}{t_{\sq} u_{\sq}}\left(1 -\!\frac{s\ms^2}
 {t_{\sq}u_{\sq}}\right)\right],
 \eea
where $C_O = N_C (N_C^2-1)$ and $C_K = (N_C^2-1)/N_C.$ The $t$-channel
contribution in $q_i\qb_j\rightarrow\sq\sqb$ depends on the gluino mass $\mg$
and does not contribute for $\tilde{t}$ production due to the negligible top quark
density in the photon. Furthermore, the gluino-exchange
$t$-channel term $\propto 1-
\delta_{ij}$ is the only term through which off-diagonal squarks can be
produced. Note that this is impossible in $e^+e^-$ annihilation.
The quark-initiated matrix elements have to be convolved with the
leading order ${\cal O}(\alpha/\alpha_s)$
quark densities in the photon. These are
fairly well constrained from deep-inelastic $e\gamma$ scattering and peak
at $x\simeq 1$.

If we include all direct, single-resolved, and double-resolved contributions
to up-type squark production in photon-photon scattering, we obtain the results
shown in Fig.\ \ref{fig:sigma_msr}. Here we have used the GRV (LO) parton
densities in the photon with a leading order value of $\alpha_s^{\rm n_f=5}
(\mu=\ms)$ and $\Lambda^{(4)}=200~{\rm MeV}$ \cite{Gluck:1992ee}. Resolved
processes (mostly photon-gluon fusion) contribute substantially only at small
squark masses (below 100 GeV), which are experimentally excluded. While the
$q\qb$ $t$-channel varies by an order of magnitude with the gluino mass, this
dependence does not show up in the total cross section. We have chosen
gluino masses between the current experimental limit of 200 GeV
\cite{Groom:2000in} and 1 TeV, where weak-scale supersymmetry starts
to become unnatural. The total cross
section is clearly dominated by the direct channel and has very little
dependence on resolved contributions or assumptions on the photon structure.


\section{Polarized Photon Spectra}
\label{sec:4}

Future lepton colliders are very likely to have a high degree of longitudinal
polarization $|\lambda_e| \leq 1/2$. Part of this polarization will
be transferred to the produced photons. The polarization state of the photon
is determined by the Stokes parameters $\xi_i,~i=1,2,3$ \cite{Landau}, where
$\sqrt{\xi_1^2+\xi_3^2}$ is the degree of linear polarization and 
\bea
 \xi_2  = \Delta f_{\gamma/e}(x) / f_{\gamma/e}(x) &~~~,~~~&
 \Delta f_{\gamma/e}(x)   = f_{\gamma/e}^+(x) - f_{\gamma/e}^-(x)
\eea
is the mean helicity. Since $\xi_1$ and $\xi_3$ generally depend on the
azimuthal angle, we will be mainly concerned with the circular polarization
parameter $\xi_2$.

The circularly polarized bremsstrahlung spectrum
\beq
 \Delta f_{\gamma/e}^{\rm brems}(x) = \frac{2\lambda_e\alpha}{2\pi}
 \left[\frac{1-(1-x)^2}{x} \ln \frac{Q^2_{\max} (1-x)}{m_e^2 x^2}
 +2 m_e^2 x^2\left(\frac{1}{Q^2_{\max}}-\frac{1-x}{m_e^2 x^2}\right) \right]
\eeq
has been derived in \cite{deFlorian:1999ge}. As in the unpolarized case
a non-logarithmic term is present which is, however, not singular for
$x\rightarrow 0$.

\begin{figure}
 \begin{center}
  \epsfig{file=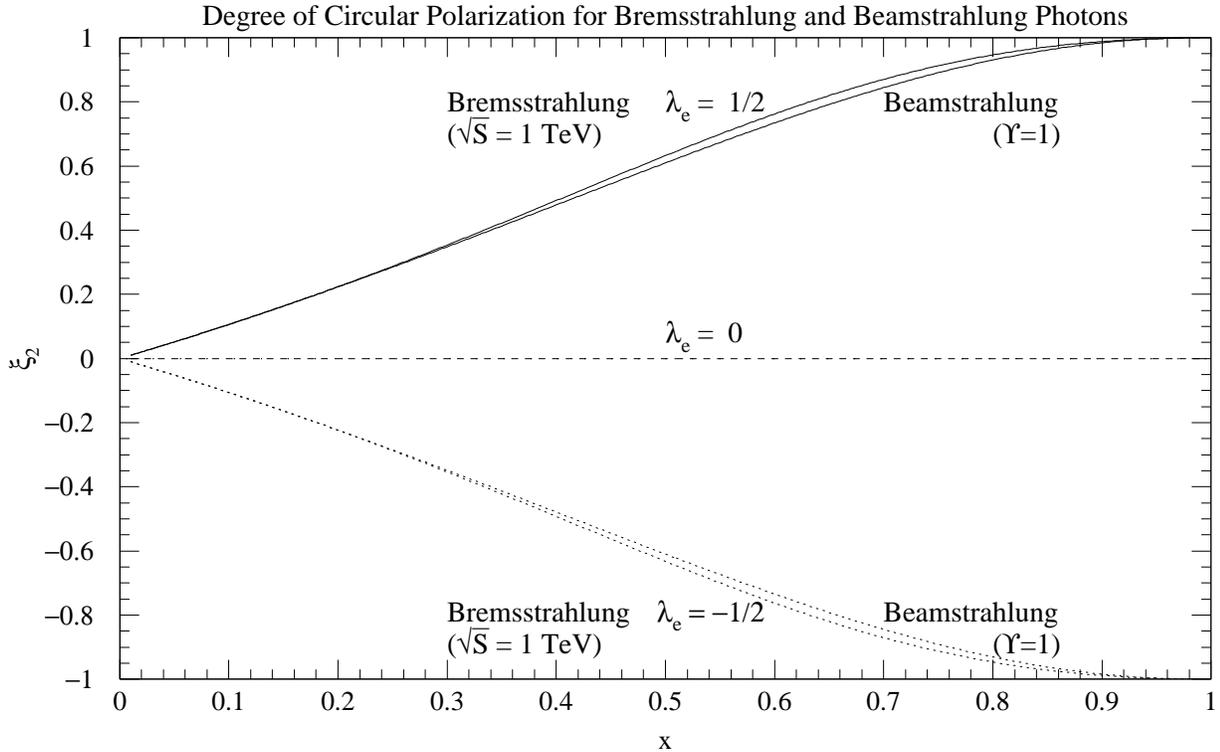,width=\textwidth,clip=}
 \end{center}
 \caption{\label{fig:brems_pol}
 Degree of circular polarization for bremsstrahlung and beamstrahlung
 photons.}
\end{figure}

\begin{figure}
 \begin{center}
  \epsfig{file=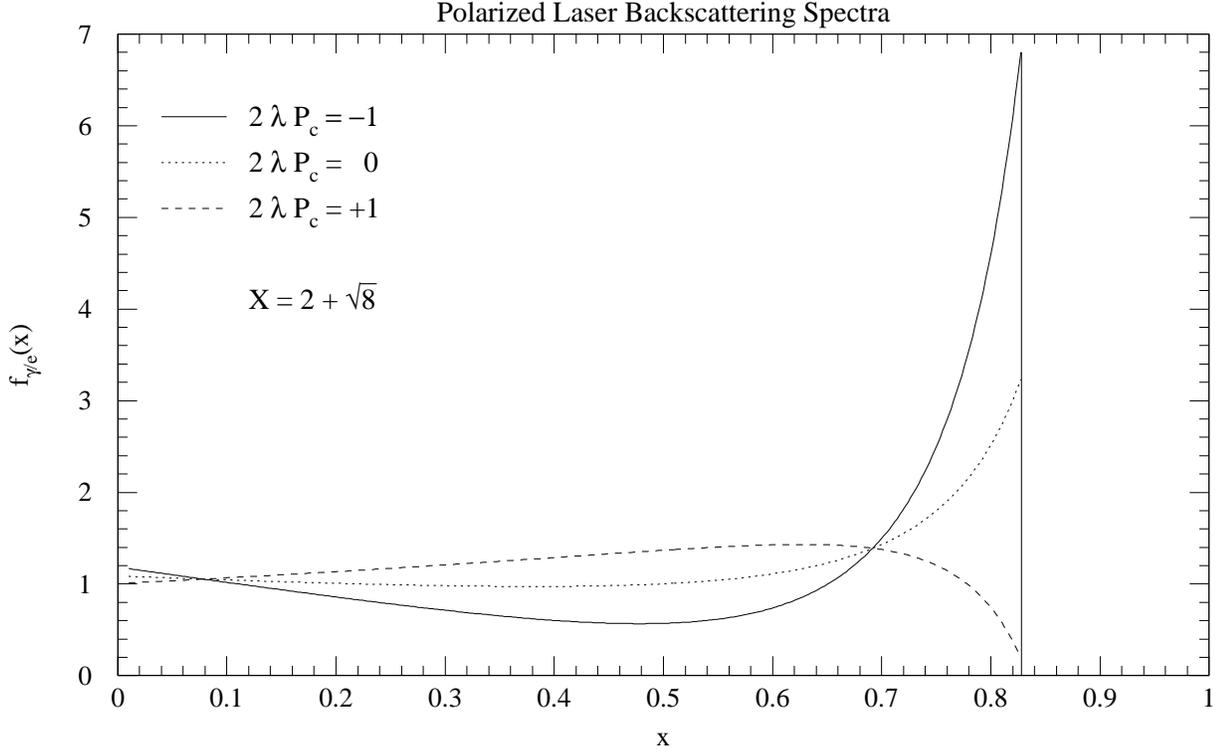,width=\textwidth,clip=}
 \end{center}
 \caption{\label{fig:phospec_pol_500}
 Laser backscattering spectra for different combinations of electron and laser
 photon polarization $2 \lambda_e P_c$.}
\end{figure}

The $x$-dependence of the circularly polarized beamstrahlung spectrum
\beq
 \Delta f_{\gamma/e}^{\rm beam}(x) = \frac{5 \lambda_e}{2\sqrt{3}\Upsilon}
 \int_u^{\infty}{\rm d}v{\rm Ai}(v)\left[\left(\frac{2v}{u}-1\right)
 \frac{1-(1-x)^2}{2(1-x)}+\frac{x^2}{2(1-x)}\right]
\eeq
is very similar, as can be seen in Fig.\ \ref{fig:brems_pol}. At $x\simeq 1$
the photons are completely polarized parallel to the incoming electron
helicity, but at $x\simeq 0$, where most of the brems- and beamstrahlung
photons are produced, they are completely unpolarized. This can be understood
from the fact that the electron polarization is lost in the Lorentz
transformation from the Breit frame of the electron-photon vertex to the
center-of-mass frame of the photon-target vertex. The dependence of $\xi_2$
on $\sqrt{S}$ and $\Upsilon$ is very weak. As $\Upsilon\rightarrow 100$,
the beamstrahlung polarization coincides with the bremsstrahlung polarization.

While the photon polarization at an $e^+e^-$ collider is thus rather limited, a
photon collider offers the additional possibility to control the helicity of
the laser photons $|P_c| \leq 1$. This also affects the unpolarized photon
spectrum \cite{Ginzburg:1984yr}
\beq
 f_{\gamma/e}^{\rm laser}(x) = \frac{1}{N_c+2 \lambda_e P_c N_c'} \left[1-x
 +\frac{1}{1-x}-\frac{4 x}{X (1-x)}+\frac{4 x^2}{X^2 (1-x)^2}
 - 2 \lambda_e P_c \frac{x (2-x) [x (X+2)-X]}{X (1-x)^2} \right] ,
\eeq
where
\beq
 N_c' = \left[\left(1+\frac{2}{X}\right)\ln(1+X)-\frac{5}{2}+\frac{1}{1+X}
 -\frac{1}{2(1+X)^2}\right]
\eeq
is related to the polarized
total Compton cross section. As can be seen in Fig.\
\ref{fig:phospec_pol_500}, the monochromaticity of the outgoing photons
can be improved considerably by choosing $2 \lambda_e P_c = -1$.

\begin{figure}
 \begin{center}
  \epsfig{file=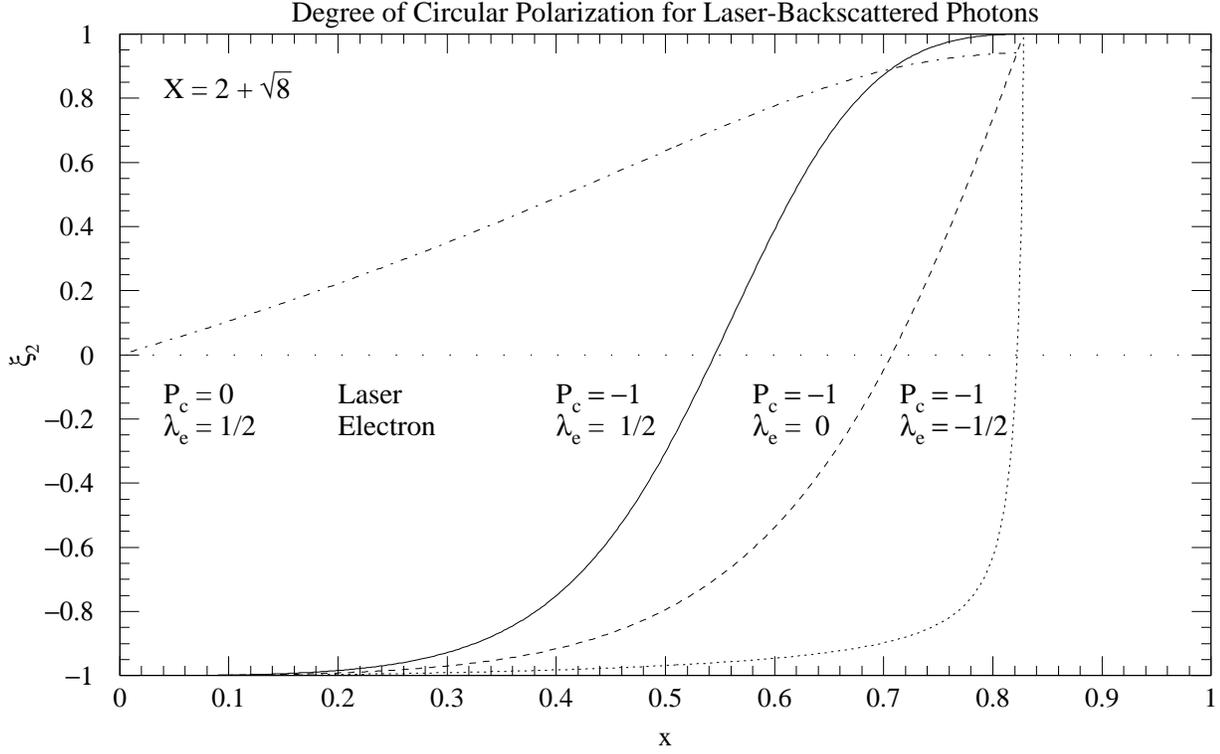,width=\textwidth,clip=}
 \end{center}
 \caption{\label{fig:laser_pol}
 Degree of circular polarization for laser-backscattered photons.}
\end{figure}

Fig.\ \ref{fig:laser_pol} shows the degree of circular polarization $\xi_2$
with
\beq
 \Delta f_{\gamma/e}^{\rm laser}(x) = \frac{1}{N_c+2 \lambda_e P_c N_c'}
 \left\{ 2 \lambda_e \frac{x}{1-x}
 \left[ 1+(1-x) \left( 1-\frac{2 x}{(1-x) X}\right)^2\right]
 + P_c \left( 1-\frac{2 x}{(1-x) X}\right) \left(1-x+\frac{1}{1-x}\right)
 \right\}
\eeq
for laser-backscattered photons \cite{Ginzburg:1984yr}.
If only the electrons are polarized $(P_c=0)$,
the result coincides again with that for brems- and beamstrahlung. However,
if $P_c = \pm 1$, then the backscattered photons have helicity $\xi_2 = -P_c$
at $x = x_{\max}$. Therefore the choice $2 \lambda_e P_c = -1$
guarantees not only good monochromaticity, but also a high degree of circular
polarization of the produced photons.
By switching the signs of $\lambda_e$ and $P_c$ simultaneously, one can
switch the helicity $\xi_2$ of the outgoing photons without changing
the photon spectrum or spoiling its monochromaticity.


\section{Polarized Cross Sections}
\label{sec:5}

The cross section for sfermion production in polarized photon-photon
collisions can again be calculated from Eqs.~(\ref{eq:ee_xsec_unpol}) and
(\ref{eq:yy_xsec_unpol}). However, the unpolarized matrix element squared
has to be replaced with
\bea
 \overline{|{\cal M}^B_{\gamma\gamma}|}^2 &=& \frac{4 e^4 e_{\sfa}^4 N_C}
 {(1-\eps)^2 t_{\sfa}^2 u_{\sfa}^2}\left\{ (1-\eps) t_{\sfa}^2 u_{\sfa}^2 
            \left[ 1+\tilde{\xi}_1^{(1)} \tilde{\xi}_1^{(2)} - 
                     \tilde{\xi}_2^{(1)} \tilde{\xi}_2^{(2)} + 
                     \tilde{\xi}_3^{(1)} \tilde{\xi}_3^{(2)} \right]\right. \\
 & & \left. - 2 \mf^2 t_{\sfa} u_{\sfa} s
            \left[ \tilde{\xi}_1^{(1)} \tilde{\xi}_1^{(2)} - 
                   \tilde{\xi}_2^{(1)} \tilde{\xi}_2^{(2)} + 
                   \left( 1 + \tilde{\xi}_3^{(1)} \right)  
                   \left( 1 + \tilde{\xi}_3^{(2)} \right)  \right]
            +  2 \mf^4 s^2 
            \left( 1 + \tilde{\xi}_3^{(1)} \right)  
            \left( 1 + \tilde{\xi}_3^{(2)} \right)
 \right\} , \nonumber
\eea
which has been calculated using the
covariant density matrix for polarized photons \cite{Landau}

\begin{figure}
 \begin{center}
  \epsfig{file=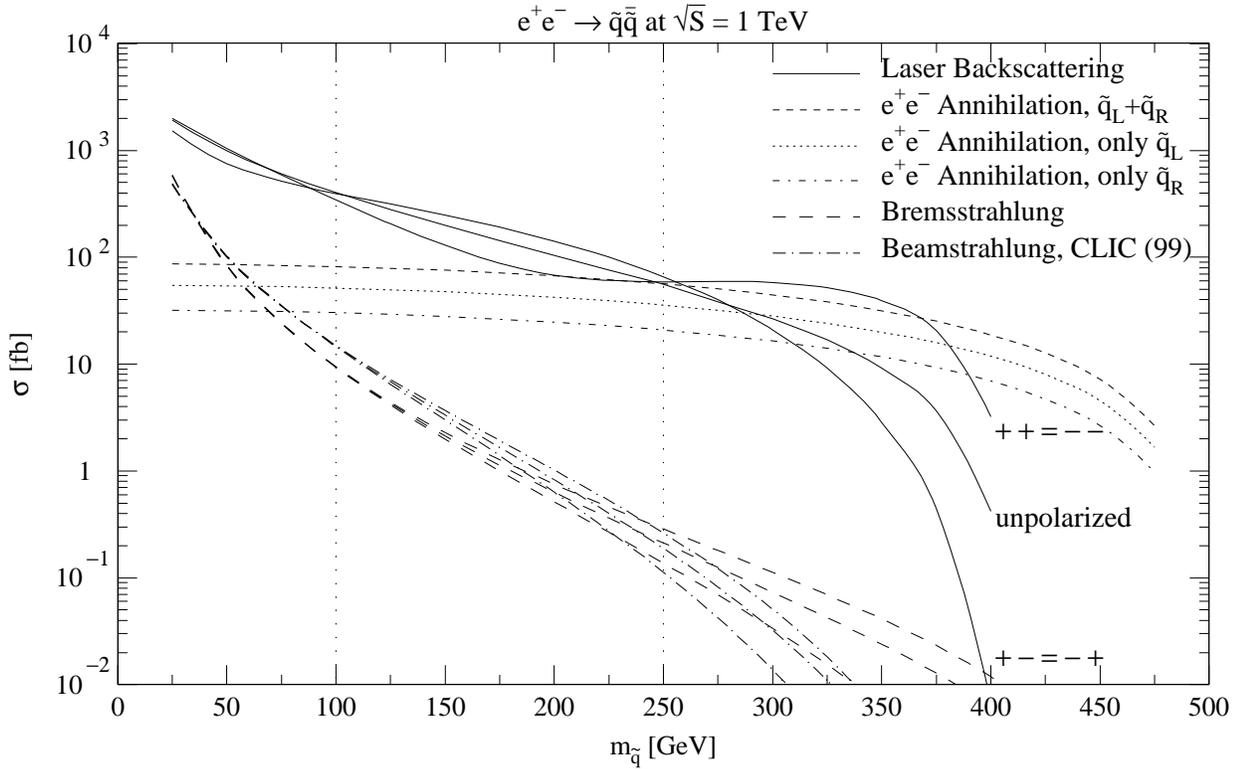,width=\textwidth,clip=}
 \end{center}
 \caption{\label{fig:sigma_ms_pol}
 Total cross sections for
 up-type squark ($\sq_L+\sq_R$) production in polarized photon-photon
 scattering and in $e^+e^-$ annihilation at a 1 TeV collider as a
 function of the squark mass.}
\end{figure}

\beq
 \rho_{\mu\nu}^{(1,2)} = \eps_\mu\eps_\nu^\ast =
 \frac{1}{2}\left(e^x_\mu e^x_\nu+
 e^y_\mu e^y_\nu\right)\pm\frac{\xi_1^{(1,2)}}{2}
 \left(e^x_\mu e^y_\nu+e^y_\mu e^x_\nu\right)
 \mp\frac{{\rm i} \xi_2^{(1,2)}}{2}
 \left(e^x_\mu e^y_\nu-e^y_\mu e^x_\nu\right)+\frac{\xi_3^{(1,2)}}{2}
 \left(e^x_\mu e^x_\nu-e^y_\mu e^y_\nu\right).
\eeq
The $\xi_i^{(1,2)},~i=1,2,3$, are the Stokes parameters discussed in the
previous Section. They describe the polarizations $\eps_{\mu,\nu}^{(\ast)}$
of the photons with momenta $k_{1,2}=\sqrt{s}/2
(1,0,0,\pm 1)$. $e^x$ and $e^y$ denote unit vectors in the $x$ and $y$
directions. The momenta of the outgoing sfermions are given by
$p_{1,2} = (m_T\cosh y,\pm p_T\cos\phi,\pm p_T\sin\phi,\pm m_T\sinh y)$.
$p_T,~y$, and $\phi$ are the transverse momentum, rapidity
and azimuthal angle of the produced sfermions, and $m_T=\sqrt{\mf^2+p_T^2}$
is the transverse sfermion mass.
The azimuthal dependence of the cross section has been included in
the rotated Stokes parameters
\bea
 \tilde\xi^{(1)}_1 &=&  \xi^{(1)}_1 \cos(2\phi)-\xi^{(1)}_3 \sin(2\phi),
 \\
 \tilde\xi^{(1)}_2 &=&  \xi^{(1)}_2,
 \nonumber\\
 \tilde\xi^{(1)}_3 &=&  \xi^{(1)}_1 \sin(2\phi)+ \xi^{(1)}_3 \cos(2\phi),
 \nonumber\\
 \tilde\xi^{(2)}_1 &=&  \xi^{(2)}_1 \cos(2\phi)+\xi^{(2)}_3 \sin(2\phi),
 \nonumber\\
 \tilde\xi^{(2)}_2 &=&  \xi^{(2)}_2,
 \nonumber\\
 \tilde\xi^{(2)}_3 &=& -\xi^{(2)}_1 \sin(2\phi)+ \xi^{(2)}_3 \cos(2\phi).
 \nonumber
\eea

For sfermion production in polarized photon-photon collisions we
consider only direct processes. Our analysis of the unpolarized cross
section has clearly demonstrated that resolved processes are only important at
very small, experimentally excluded, squark masses. Furthermore, almost
nothing is known experimentally about polarized parton densities in the photon.
Predictions for polarized resolved photoproduction would thus be very
speculative.

\begin{figure}
 \begin{center}
  \epsfig{file=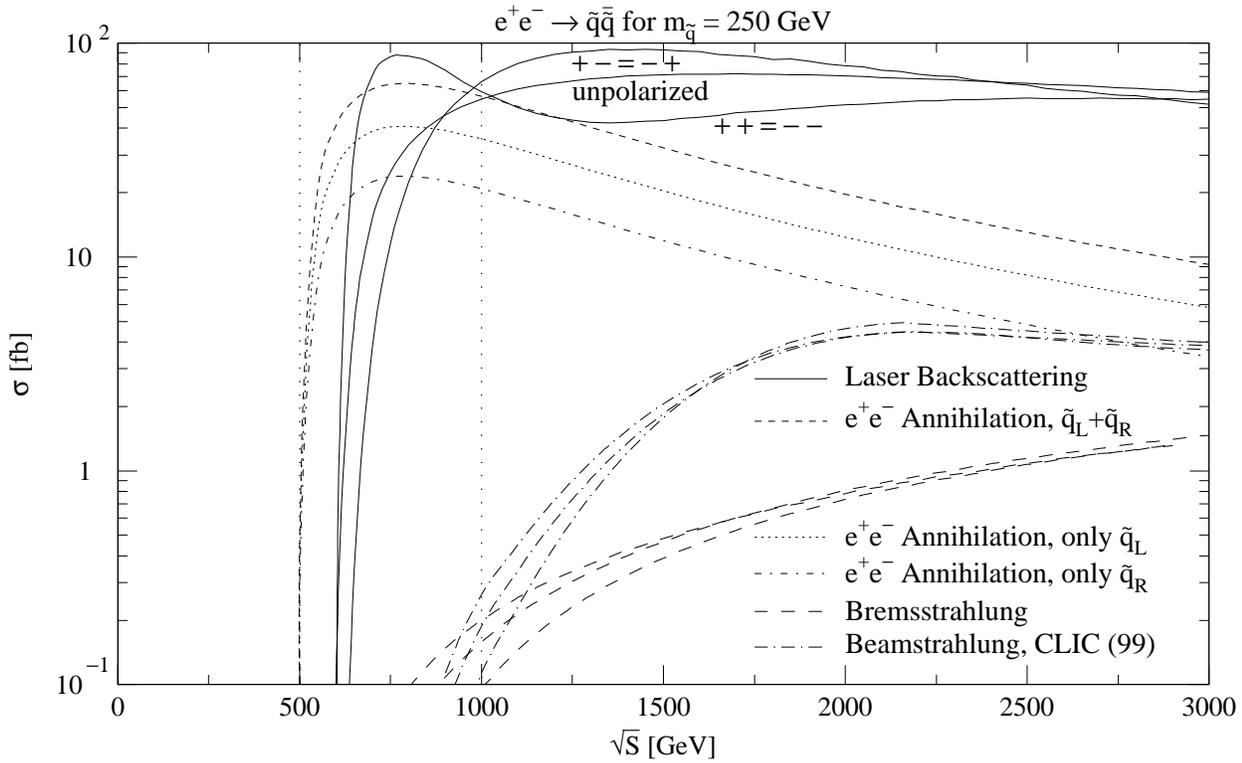,width=\textwidth,clip=}
 \end{center}
 \caption{\label{fig:sigma_250_s_pol}
 Total cross sections for the production of
 up-type squarks ($\sq_L+\sq_R$) of mass 250 GeV in
 polarized photon-photon scattering and in $e^+e^-$ annihilation as a
 function of the center-of-mass energy.}
\end{figure}

Because of the azimuthal dependence, cross sections with linear photon
polarization are difficult to disentangle and remain small if averaged over
the azimuthal angle. Therefore we restrict ourselves to circularly polarized
photons and set $\xi_1=\xi_3=0$. Since
$\overline{|{\cal M}^B_{\gamma\gamma}|}^2$ depends then only the product
$\tilde{\xi}_2^{(1)} \tilde{\xi}_2^{(2)}$, we expect identical cross sections
for incoming photons with identical or opposite helicities.

In Fig.\ \ref{fig:sigma_ms_pol} we compare the $e^+e^-$ annihilation
cross section against the polarized photon-photon cross section for
up-type squark production at a 1 TeV photon collider for different squark masses.
The labels $++,~--,~+-$, and $-+$ denote the helicities $P_c$
of the incoming laser photons. The helicities of the incoming leptons
$\lambda_e$ have
always been chosen to ensure the condition for optimal monochromaticity,
$2\lambda_e P_c=-1$. The backscattered photons are therefore highly polarized
in the direction opposite to the laser photon (see Fig.\ \ref{fig:laser_pol}).
The unpolarized curve is the
same as in Fig.\ \ref{fig:sigma_ms}, {\it i.e.} $\lambda_e=P_c=0$.

Fig.\ \ref{fig:sigma_ms_pol} demonstrates that the unpolarized photon-photon
cross section can be enhanced in the region $\mf \in [100;250]$ GeV by about
40\% if one chooses opposite laser photon helicities ($+-$ or $-+$). For
$\mf > 250$ GeV the effect is even more dramatic: The cross section can be
improved by almost an order of magnitude at large $\mf$ if one chooses
identical laser photon helicities. The cross section at a polarized photon
collider stays larger than the $e^+e^-$ annihilation cross section almost up
to the kinematic limit of the photon collider. It is interesting to note that
one has to switch from opposite to identical helicities at $\mf \simeq 250$
GeV, where the unpolarized photon-photon cross section drops below the
annihilation cross section.
In Fig.\ \ref{fig:sigma_ms_pol} we also show polarization effects for sfermions
produced via brems- and beamstrahlung. The cross sections remain small and are
only slightly enhanced by preferring identical over opposite lepton helicities.

In Fig.\ \ref{fig:sigma_250_s_pol} we compare the same cross sections
for a fixed up-type squark mass of 250 GeV as a function of the center-of-mass
energy of the collider $\sqrt{S}$. The unpolarized photon cross section can
again be enhanced by an appropriate choice of polarization. In particular,
identical laser photon helicities lead to a photon cross section which is
larger than the annihilation cross section already at the threshold of the
photon collider, {\it i.e.} below 1 TeV, while the cross section for
opposite helicities is smaller by a factor $\beta^4=(1-4\ms^2/s)^2$.
Polarization for brems- and beamstrahlung is 
\begin{figure}
 \begin{center}
  \epsfig{file=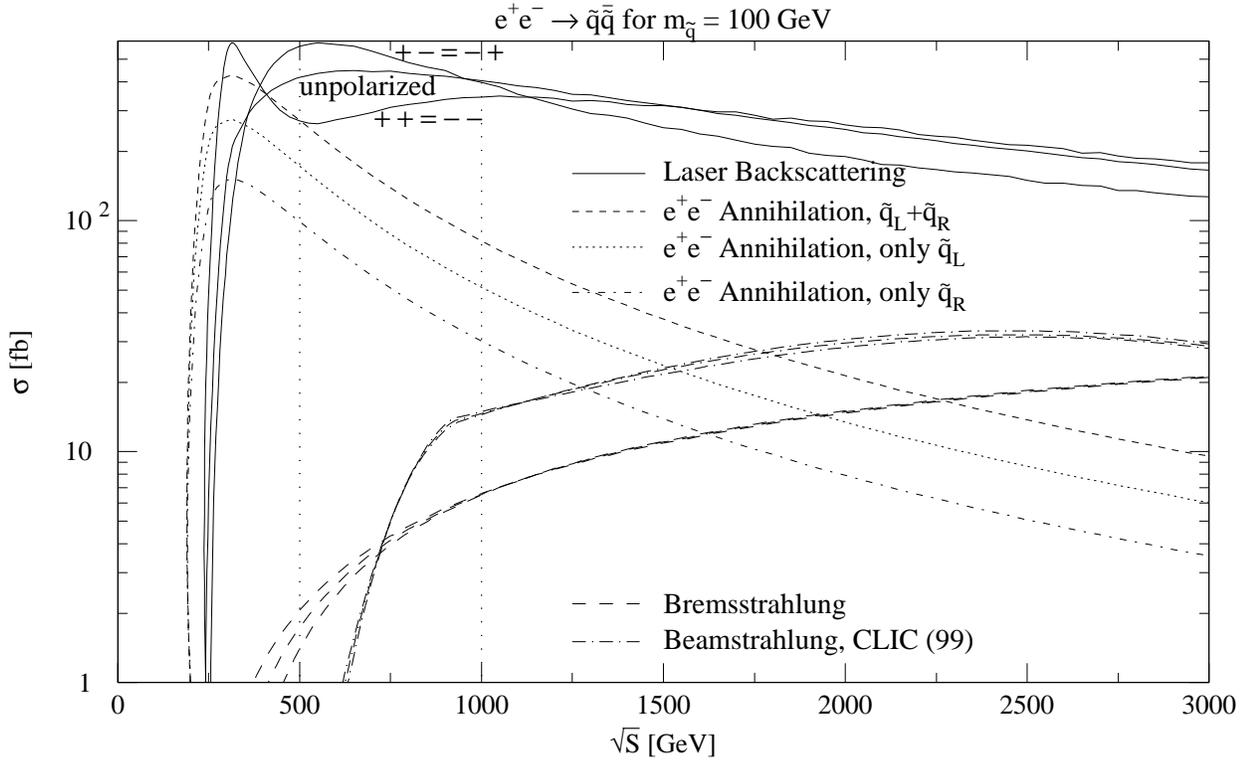,width=\textwidth,clip=}
 \end{center}
 \vspace*{-8mm}
 \caption{\label{fig:sigma_100_s_pol}
 Total cross sections for the production of
 up-type squarks ($\sq_L+\sq_R$) of mass 100 GeV in
 polarized photon-photon scattering and in $e^+e^-$ annihilation as a
 function of the center-of-mass energy.}
\end{figure}
\vspace*{-16mm}
\begin{figure}
 \begin{center}
  \epsfig{file=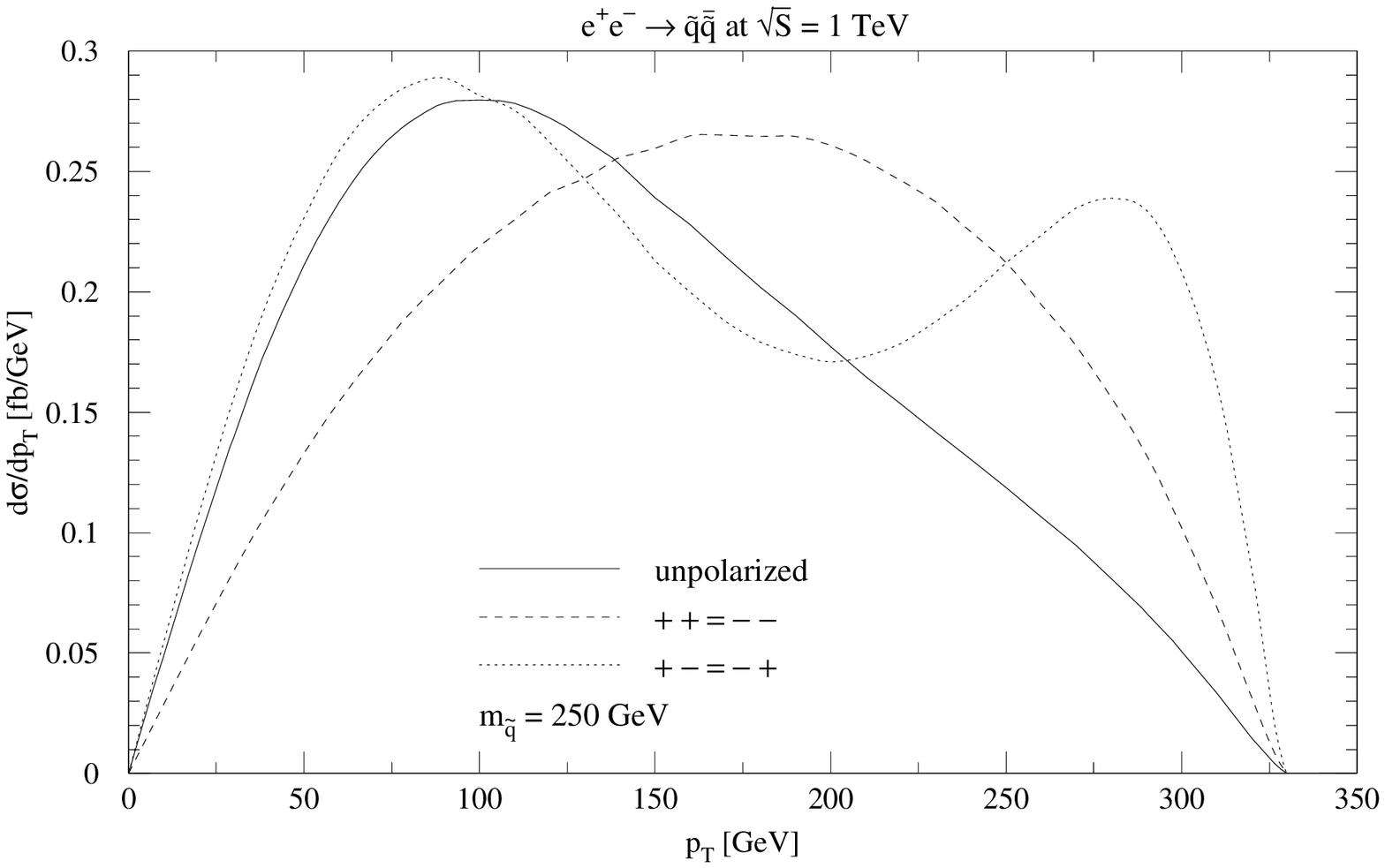,width=\textwidth,clip=}
 \end{center}
 \vspace*{-8mm}
 \caption{\label{fig:sigma_1tev_pt}
 Differential cross sections for the production of
 up-type squarks ($\sq_L+\sq_R$) of mass
 250 GeV at a 1 TeV polarized photon collider as a function of the
 transverse momentum $p_T$.}
\end{figure}
\begin{figure}
 \begin{center}
  \epsfig{file=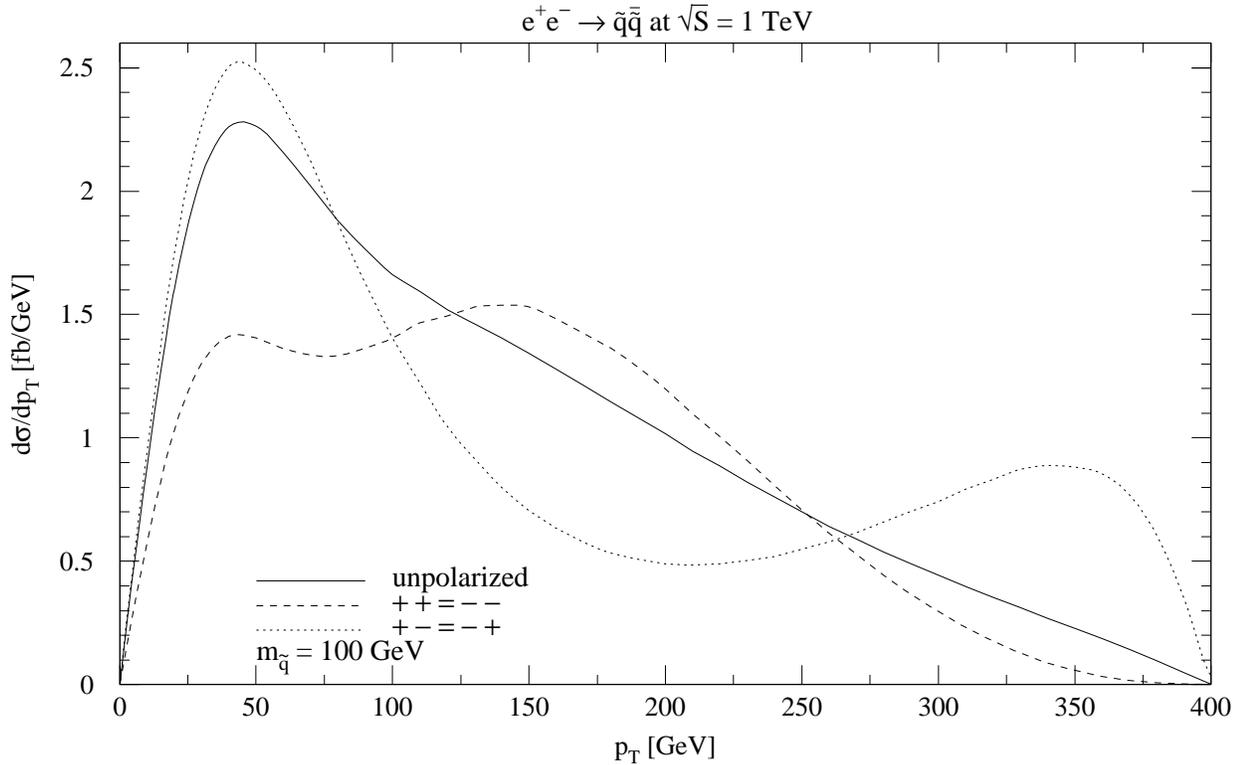,width=\textwidth,clip=}
 \end{center}
 \caption{\label{fig:sigma_1tev_100gev_pt}
 Differential cross sections for the production of
 up-type squarks ($\sq_L+\sq_R$) of mass
 100 GeV at a 1 TeV polarized photon collider as a function of the
 transverse momentum $p_T$.}
\end{figure}
\noindent
again of little interest. At large
$\sqrt{S}$, where the cross sections become large, the radiated
photons are completely unpolarized.

For up-type squarks of mass 100 GeV we show the center-of-mass energy
dependence in Fig.\ \ref{fig:sigma_100_s_pol}. Again the cross section
at threshold can be optimized by choosing identical photon helicities.

For experimental analyses it is also important to study differential cross
sections, {\it e.g.} in the transverse momentum $p_T$ or the
rapidity $y$ of the produced supersymmetric particles,
\beq
 \frac{{\rm d}\sigma_{e^+e^-}^B(S)}{{\rm d}p_T{\rm d}y}
 = 2 p_T S \int {\rm d}x_1 x_1 f_{i/e}(x_1) {\rm d}x_2 x_2 f_{j/e}(x_2)
 \frac{{\rm d}^2\sigma^B_{ij}(s)}
 {{\rm d}t_{\sfa}\,{\rm d}u_{\sfa}},
\eeq
since cuts on $p_T$ and $y$ can help to eliminate backgrounds.
For this reason we show in Fig.\
\ref{fig:sigma_1tev_pt} differential $p_T$ spectra for up-type squarks
of mass 250 GeV, produced at a 1 TeV photon collider. The spectra have
been integrated over the rapidity $y$ and extend
out to the kinematic limit $p_T < 0.828 \sqrt{S} - 2 \mf = 328$ GeV.
While the unpolarized spectrum peaks at $p_T \simeq 100$ GeV, or roughly
at $\mf/2$, the mean $p_T$ of sfermions produced with backscattered laser
photons of identical helicity is almost twice as big. If the laser photons have
opposite helicity, one gets a distinct twin-peak behavior with a local minimum.
This is due to the absence of the four-point interaction diagram in Fig.\
\ref{fig:dlo} which contributes only for identical helicities at intermediate
$p_T$. These features should be very helpful in experimental analyses.
Similar results for up-type squarks of mass 100 GeV are shown in Fig.\
\ref{fig:sigma_1tev_100gev_pt}.

\begin{figure}
 \begin{center}
  \epsfig{file=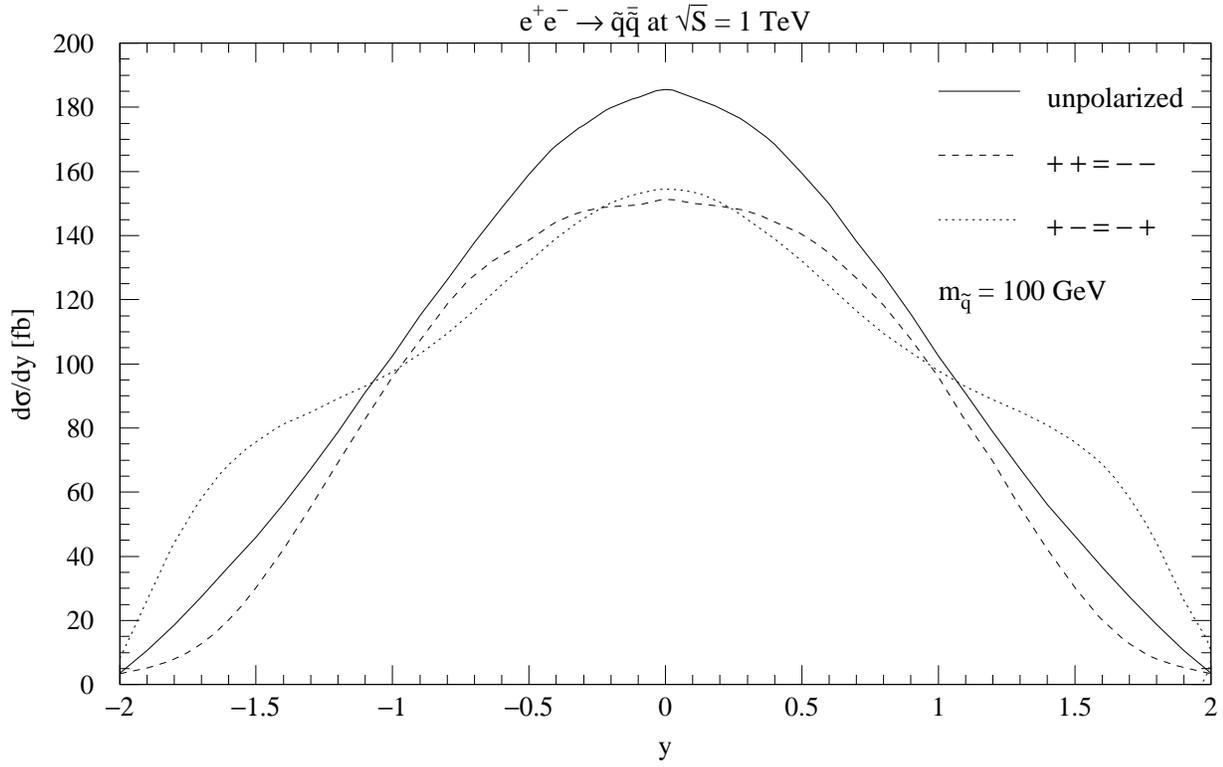,width=\textwidth,clip=}
 \end{center}
 \vspace*{-8mm}
 \caption{\label{fig:sigma_1tev_y2}
 Differential cross sections for the production of
 up-type squarks ($\sq_L+\sq_R$) of mass
 100 GeV at a 1 TeV polarized photon collider as a function of the
 rapidity $y$.}
\end{figure}

\begin{figure}
 \begin{center}
  \epsfig{file=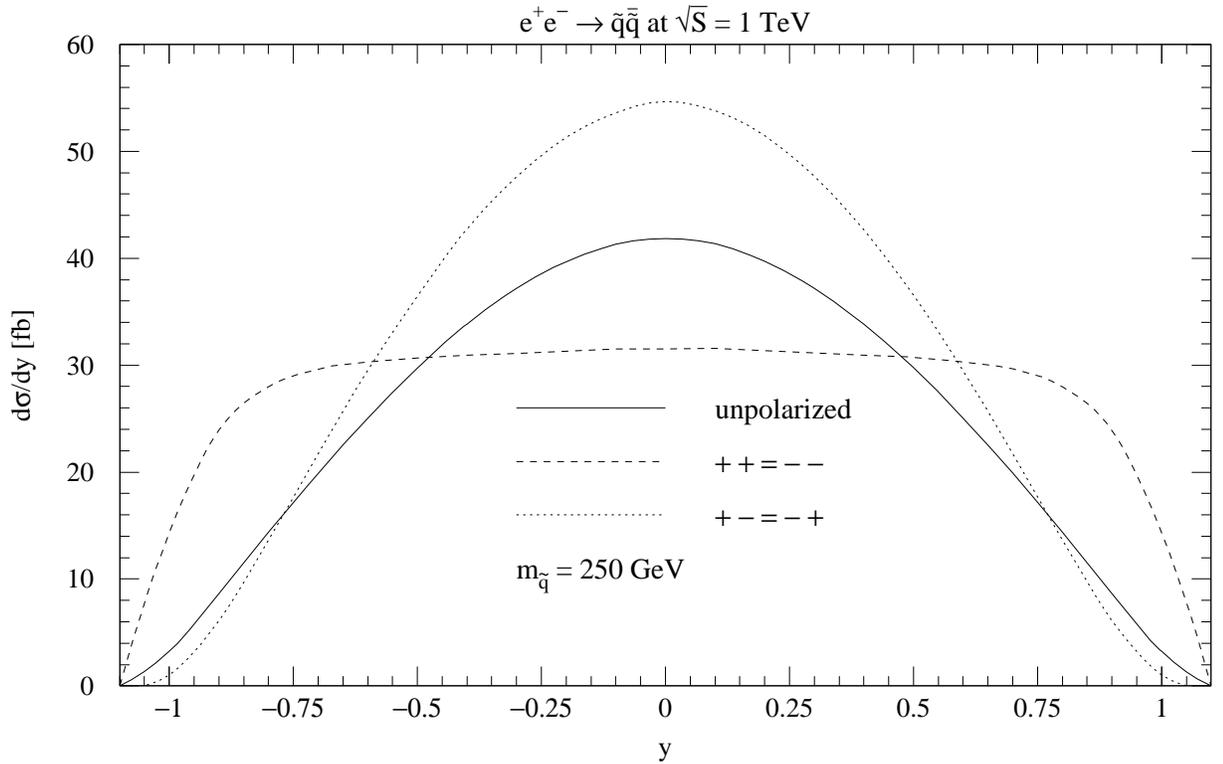,width=\textwidth,clip=}
 \end{center}
 \vspace*{-8mm}
 \caption{\label{fig:sigma_1tev2_y2}
 Differential cross sections for the production of
 up-type squarks ($\sq_L+\sq_R$) of mass
 250 GeV at a 1 TeV polarized photon collider as a function of the
 rapidity $y$.}
\end{figure}

Finally we show in Fig.\ \ref{fig:sigma_1tev_y2} rapidity distributions
for up-type squarks of mass 100 GeV produced at a 1 TeV photon collider.
The rapidity spectra are symmetric at  $y=0$ and extend out to $|y| < 2$.
This range should be covered by the detector at a photon collider to provide
optimal analyzing conditions for sfermions of mass $\mf = 100$ GeV. For
$\mf=250$ GeV the rapidity spectrum is narrower and extends out to
$|y| < 1.1$ (see Fig.\ \ref{fig:sigma_1tev2_y2}).
The spectrum for laser photons with identical helicities is again very similar
to the unpolarized spectrum. The spectrum for opposite helicities has
interesting shoulders at $y= \pm 1.5$ in the case of $\mf=100$ GeV.
The dips at $y=\pm 1$
are again due to the absence of the four-point interaction diagram in Fig.\
\ref{fig:dlo} for opposite helicities.

\vspace*{-16mm}


\section{Conclusion}
\label{sec:6}

In this paper we have presented a detailed analysis of sfermion production 
in photon-photon collisions. We have reviewed the unpolarized and polarized
photon spectra coming from bremsstrahlung, beamstrahlung, and laser
backscattering and updated them using the latest linear collider design
parameters. We have calculated for the first time
total and differential cross sections for sfermion production in photon-photon
collisions, including contributions from resolved photons and arbitrary photon
polarization. Our numerical results have been compared directly to the
competing $e^+e^-$ annihilation cross section.
We have chosen to present our results for the typical case of up-type squark
production. The cross sections for down-type squarks and sleptons can easily
be obtained by rescaling our results according to the sfermion charge
and color factor.

Brems- and beamstrahlung photons will be produced naturally at any linear
$e^+e^-$ collider. Our results show that the corresponding production
cross sections are small, except at the very large center-of-mass energies
envisaged in the CLIC design. The polarization of the initial photons remains
small even if the initial lepton polarization is large. Therefore polarization
effects in brems- and beamstrahlung are of little interest.

A dedicated photon collider will require the construction of additional laser
facilities at some extra cost. However, we have demonstrated that a photon
collider may be advantageous for the analysis of sfermions for several
reasons:

\begin{itemize}

\item In leading order of perturbation theory, photon-photon collisions are
pure SUSY-QED processes, which depend only on the physical sfermion mass and
not on the details of the SUSY breaking mechanism. Any model dependence can
therefore be analyzed cleanly in the decay of the sfermions.

\item The photon cross section is very sensitive to the sfermion charge so
that sleptons, up-type, and down-type squarks can be clearly distinguished.

\item A photon collider can produce almost monochromatic photons, {\it i.e.}
photons which have about 83 \% of the electron beam energy, and they can be
highly polarized. As a consequence we find that the production cross
sections are larger than those in $e^+e^-$ annihilation for a large range of
sfermion masses. If the incoming laser photons and leptons are polarized and
the photons have identical helicities, this is
even true up to the kinematic limit of the photon collider. For lower sfermion
masses or higher center-of-mass energies the cross section can be improved by
about 40 \% by choosing opposite laser photon helicities.

\end{itemize}

Resolved photon processes are only important for squarks, since sleptons do
not couple strongly. Furthermore, resolved photons contribute significantly
only at very small squark masses, which are experimentally already excluded.
While there is thus no enhancement of the production cross section, there is
also no uncertainty due to the photon structure or the scale of the strong
coupling constant.

Differential cross sections are important for experimental analyses to
distinguish signal from background events. We find that the $p_T$ spectrum
for sfermion production in photon collisions peaks roughly at $\mf/2$ as
expected. The rapidity spectrum for sfermions of mass $\mf=100$ GeV at a 1 TeV
photon collider extends out
to $\pm 2$. We conclude that this should be the minimum coverage a detector at
a photon collider should have to
ensure full analyzing power. The polarized $p_T$ and $y$ spectra have very
distinct features which should be helpful in the experimental analysis.


\section*{Acknowledgments}

We thank D.~V.~Schroeder for mailing us a copy of his Ph.~D.\ thesis and
K.~Hagiwara,
B.~A.~Kniehl, G.~Kramer, V.~Telnov, and A.~Tkabladze for valuable comments.
This work has been supported by the Deutsche Forschungsgemeinschaft through
Grant No.\ KL~1266/1-1 and Graduiertenkolleg {\it Zuk\"unftige Entwicklungen
in der Teilchenphysik} and by the European Commission through Grant
No.\ ERBFMRX-CT98-0194.






\end{document}